\documentclass[aps,physrev,superscriptaddress]{revtex4-2}

\usepackage{graphicx}
\usepackage{xcolor}
\usepackage{bbold}
\usepackage{amsmath}
\usepackage{amssymb}
\usepackage{comment}

\usepackage{multirow} 
\usepackage{pifont}
\usepackage{verbatim}
\usepackage{booktabs}
\usepackage{esint}
\usepackage{bm}
\usepackage[mathlines]{lineno}
\usepackage{makecell}

\newcommand{\fomega}{\overline{\omega}}
\newcommand{\fpsi}{\overline{\psi}}
\newcommand{\vel}{\bm{u}}
\newcommand{\fvel}{\overline{\vel}}
\newcommand{\average}[1]{\langle{#1}\rangle}

\begin{document}
\raggedbottom

\title{\textbf{Modified dynamic mixed subgrid-scale models \\ for geophysical flows: \\Forced two-dimensional and beta-plane turbulence
}
}

\author{Anantha Narayanan Suresh Babu}
\affiliation{Department of Mechanical Engineering, Massachusetts Institute of Technology, Cambridge, USA}
\affiliation{Center for Computational Science and Engineering, Massachusetts Institute of Technology, Cambridge, USA}
\author{Akhil Sadam}
\affiliation{Center for Computational Science and Engineering, Massachusetts Institute of Technology, Cambridge, USA}
\author{Pierre F.J. Lermusiaux}
\email[Contact author: ]{pierrel@mit.edu}
\affiliation{Department of Mechanical Engineering, Massachusetts Institute of Technology, Cambridge, USA}
\affiliation{Center for Computational Science and Engineering, Massachusetts Institute of Technology, Cambridge, USA}

\date{\today}

\begin{abstract} 
Subgrid-scale (SGS) models for large-eddy simulations (LES) of geophysical turbulence typically need to balance dissipative regularization with backscatter, the upscale transfer of energy from unresolved to resolved scales. 
Dynamic mixed models (DMMs) combine functional eddy viscosity and structural closures through dynamically estimated coefficients that are least-squares optimal with respect to the Germano identity error (GIE). We show that this classical DMM least-squares estimation can be dominated by the structural component, thereby limiting the functional component's dissipative regularizing role. To address this limitation, we decompose the Gram matrix formed from the inner products of the functional and structural contributions to the GIE based on their self- and cross-interactions. We then develop a modified Gram-based framework to construct a novel parametric family of fully-coupled, sequential, and fully-decoupled DMMs with tunable structural-functional balance. Using an idealized forced two-dimensional mesoscale and beta-plane turbulence framework with
the Leith model as the functional component and the
fourth-order nonlinear gradient model as the structural component, we evaluate the resulting closures \emph{a priori} and \emph{a posteriori} across eddy ($\beta = 0$) and jet ($\beta > 0$) regimes at two LES resolutions. The \emph{a priori} results show that structurally-dominated models achieve strong agreement with the \emph{ideal} SGS forcing
and accurately reproduce local SGS energy exchange, including backscatter.
However, these instantaneous improvements do not directly translate into long-term LES accuracy: in \emph{a posteriori} tests, structurally dominated models exhibit noise-like artifacts in the vorticity fields with high-wavenumber spectral deviations, indicating insufficient net dissipation.
In contrast, the sequential DMM in which the functional component is determined first and then corrected by the structural component retains much of the \emph{a priori} 
structural accuracy while improving the \emph{a posteriori} vorticity fields, spectra, and domain-averaged diagnostics. 
Spectral SGS energy and enstrophy-transfer analyses show
that this sequential DMM permits backscatter at scales larger than the forcing scale with enhanced dissipation at smaller scales, thereby improving the balance between instantaneous structural fidelity and long-term accuracy.
\end{abstract}


\maketitle

\section{\label{sec:introduction} Introduction}

Turbulence is a defining feature of geophysical fluid dynamics, including for larger-scale atmospheric and oceanic flows \cite{pedlosky2013geophysical, vallis2017atmospheric}. Characterized by their thin aspect ratio, stable density stratification, and planetary rotation, these flows typically exhibit two-dimensional or quasi-two-dimensional behavior \cite{boffetta2012two, danilov2000quasi}. Typically driven by large-scale forcing, such as wind stress for the ocean, forced two-dimensional (2D) turbulence differs fundamentally from its three-dimensional (3D) counterpart. While 3D turbulence exhibits a direct cascade of energy from large scales to small scales, forced 2D turbulence shows a dual cascade: an inverse cascade of energy from small to large scales accompanied by a forward cascade of enstrophy \cite{kraichnan1967inertial}. Observational evidence for the inverse cascade in oceanic turbulence has been obtained from satellite altimetry \cite{scott2005direct}. This inverse cascade is ultimately arrested by large-scale dissipative mechanisms such as Ekman friction or bottom drag, preventing the pile-up of energy \cite{smith2002turbulent}. Planetary rotation further forces the dynamics, leading to the formation of anisotropic features such as Rossby waves, zonal jets, and ocean striations \cite{srinivasan2012zonostrophic,galperin2019zonal}. 

Direct numerical simulation (DNS) of geophysical turbulence across its vast range of spatial (from Kolmogorov to basin-scale) and associated temporal scales is computationally intractable, and hence parameterizations of unresolved scales, variables, and processes are required \cite{warner2010numerical, balaji2021climbing}. In large eddy simulation (LES) models, large-scale motions are explicitly resolved while the effects of unresolved fine scales are modeled through a subgrid-scale (SGS) closure term \cite{galperin1993large}. SGS models can be broadly classified into structural models, which directly approximate the form of the SGS forcing, and functional models that do not approximate the form but rather the interactions (e.g., energy transfer) between resolved and unresolved scales \cite{sagaut2006large}. In 2D geophysical turbulence, this distinction is particularly significant because inter-scale transfers are not purely dissipative and may involve substantial backscatter, the local transfer of energy from unresolved to resolved scales \cite{piomelli1991subgrid}.

Functional closures based on eddy viscosity parameterizations, such as the Leith model, have shown promise in mesoscale LES of the ocean \cite{leith1996stochastic, fox2008can}. However, these models are typically purely dissipative and do not account for backscatter, leading to spurious energy dissipation. Dynamic functional closures partially alleviate this limitation by allowing spatiotemporal variability in the eddy viscosity \cite{germano1991dynamic}. Yet unconstrained dynamic formulations can introduce numerical instabilities, and practical implementations often rely on spatial averaging to limit backscatter \cite{ghosal1995dynamic}. In geophysical applications, backscatter is often incorporated through additional negative viscosities or stochastic parameterizations that selectively re-inject energy \cite{frederiksen2013subgrid}.
Structural models offer a more natural approach to incorporate backscatter, as they exhibit stronger correlation with the SGS forcing \cite{horiuti1989role, meneveau2000scale}. The tensor-diffusivity or nonlinear gradient model \cite{leonard1975energy, clark1979evaluation}, for example, possesses certain conservative properties, but can lead to numerical instabilities due to enstrophy growth \cite{nadiga2011equivalence}.

Mixed models offer a hybrid approach by combining structural and functional models \cite{bardina1983improved}. In dynamic mixed formulations, the relative contributions of the individual components evolve algebraically \cite{sarghini1999scale, sagaut2000general}. However, it has been observed that the structural component often dominates due to its stronger correlation with the exact SGS forcing \cite{salvetti1995priori}. As a result, the dissipative control provided by the functional term can become insufficient, potentially leading to long-term instabilities and inaccuracies. 
Though various modifications to the dynamic procedure for functional and mixed models have been proposed \cite{morinishi2001recommended, agrawal2022non, iyer2024efficient, hu2024hidden}, their impact on energy exchange and backscatter has received limited attention \cite{wang2005dynamic, shi2008constrained, vollant2016dynamic, perezhogin2023subgrid}. 
Hence, some open questions remain: Can the dynamic procedure be systematically modified to generate a controllable family of mixed models? 
How does the relative balance between structural and functional closures affect backscatter and inter-scale energy and enstrophy transfers? And how do these choices influence the trade-off between  \emph{a priori} (instantaneous) fidelity and \emph{a posteriori} (time-integrated) stability and accuracy?
And how do they affect spectral and probabilistic accuracies (e.g., peaks, bulk, and tails)? 

Previous studies have evaluated various functional models and their dynamic variants in idealized geophysical turbulence \cite{ozgokmen2007large, graham2013framework, matharu2022optimal, frezat2022posteriori, suresh_babu_et_al_Oceans2025} and more realistic ocean flows \cite{ringler2013multi, pearson2017evaluation, wilder2025examining}. However, the behavior of mixed models and their dynamic variants, 
particularly their influence on backscatter and spectral energy and enstrophy transfers, has not been thoroughly characterized.
Recent data-driven SGS closures trained \emph{a priori} (off-line) on high-resolution simulations effectively act as structural closures, as they aim to predict the exact SGS forcing \cite{ross2023benchmarking, srinivasan2024turbulence}. Such approaches may therefore inherit similar stability and backscatter characteristics. Hybrid data-driven mixed models have also been proposed \cite{kang2023neural, shankar2025differentiable}, and interpretable closed-form expressions resembling the nonlinear gradient model have been discovered for geophysical flows \cite{zanna2020data, jakhar2024learning, jakhar2026analytical}. 
Understanding the behavior of structural and mixed models is therefore crucial 
for interpreting and designing machine-learning-based closures \cite{ahmed2021closures, choi2025perspective}.

In this work, we examine how the coupling between structural and functional components in dynamic mixed models affects the trade-off between \emph{a priori} accuracy and \emph{a posteriori} performance. We analyze the standard dynamic least-squares formulation and show that classical fully-coupled dynamic procedures can become structurally dominated,
limiting the regularizing dissipative role of the functional component. To control this coupling and develop modified mixed models, we reformulate the dynamic procedure by decomposing the associated Gram matrix formed from the inner products of the functional and structural components. We then construct a parameterized family of fully-coupled, sequential, and fully-decoupled dynamic mixed models with varying levels of structural and functional contributions. Using an idealized forced mesoscale beta-plane framework, we combine the Leith subgrid-viscosity model with the fourth-order nonlinear gradient model and evaluate the resulting closures \emph{a priori} and \emph{a posteriori} in terms of SGS structural fidelity, inter-scale energy and enstrophy transfers in physical and spectral space, and the simulated vorticity fields and their non-Gaussian distributions.

This paper is organized as follows. Sec.\;\ref{sec:gov_eq} provides the governing equations and subgrid-scale contributions to the resolved kinetic energy and enstrophy. We revisit classical SGS models in Sec.\;\ref{sec:classical_sgs}, and motivate and introduce our modified dynamic mixed modeling framework in Sec.\;\ref{sec:methods}. Details of the numerical solver and test cases are presented in Sec.\;\ref{sec:test_cases}. In Sec.\;\ref{sec:results}, we perform detailed \emph{a priori} and \emph{a posteriori} tests, and present the results and discussion. Sec.\;\ref{sec:conclusions} summarizes the main findings and conclusions. 
\section{\label{sec:gov_eq} Governing Equations and Subgrid-Scale Dynamics}

We consider the one-layer 2D incompressible Quasi-Geostrophic (QG) equations in vorticity-streamfunction ($\omega$, $\psi$) form \cite{vallis2017atmospheric, suresh_babu_et_al_JAMES2026},
\begin{eqnarray}\label{eq:QG}
\begin{aligned}
    \frac{\partial \omega}{\partial t} + J(\psi,\omega) & = \nu \nabla^2 \omega - \mu \omega - \beta \psi_x + F, \\
     \omega & = \nabla^2 \psi, 
 \end{aligned}
\end{eqnarray} 
where $J(\psi,\omega) = \psi_x \omega_y - \psi_y \omega_x$ is the nonlinear advection term, $\nu$ is the viscosity, $\mu$ the linear frictional damping (bottom drag) coefficient, $\beta$ the Rossby parameter, and $F$ the external (wind) forcing. Velocity 
can be obtained from the streamfunction as $\vel = (- \psi_y, \psi_x)$, where the subscripts
$\bullet_x$ and $\bullet_y$ denote spatial derivatives.

The corresponding Large Eddy Simulation (LES) equations \citep{sagaut2006large} are obtained by applying a filtering operator $\overline{(\bullet)}$ to \eqref{eq:QG}. Filtering can be implicit or explicit \cite{kent2012assessing, volpiani2024comprehensive, quaini2024bridging}. 
In this work, we use homogeneous explicit filters such that the filtering operator and the linear derivative operators commute (the specific filter forms and parameters used are provided in Appendix\;\ref{sec:AppA}). However, the nonlinear advection term does not commute with filtering, yielding,
\begin{eqnarray}\label{eq:QG_LES}
\begin{aligned}
    \frac{\partial \fomega}{\partial t} + J(\fpsi,\fomega) & = \nu \nabla^2 \fomega - \mu \fomega - \beta \fpsi_x + \overline{F} + \Pi, \\
     \fomega & = \nabla^2 \fpsi. 
 \end{aligned}
\end{eqnarray} 
where the \emph{ideal} subgrid-scale (SGS) forcing, $\Pi = J(\fpsi,\fomega) - \overline{J(\psi,\omega)}$, represents the effect of the unresolved scales (subgrid-scales) on the resolved scales. 
``\emph{Ideal}'' indicates that this SGS forcing contains contributions from unresolved variables, which are not available in LES. An alternative flux formulation can be used for nonlinear advection and SGS contributions, $J = \nabla \cdot (\vel\,\omega)$ and $\Pi = \nabla \cdot \bm{\sigma}$. The \emph{ideal} SGS vorticity flux then corresponds to
$\bm{\sigma} = \fvel\,\fomega -\overline{\vel\,\omega}$.  

In SGS closure, $\Pi$ is modeled solely from the resolved variables $(\fomega, \fpsi)$, but must accurately represent contributions to the resolved kinetic energy $(\overline{E})$ and enstrophy $(\overline{Z})$, 
\begin{equation}\label{eq:2D_EZP}
     E  =  \frac{1}{2} \langle  |\nabla{\psi}|^2 \rangle,  \qquad
     \overline{E}  =  \frac{1}{2} \langle  |\nabla{\fpsi}|^2 \rangle, \qquad
     Z  =  \frac{1}{2} \langle  \omega^2  \rangle, \qquad
     \overline{Z}  =  \frac{1}{2} \langle  \fomega^2  \rangle,
\end{equation}
where $\langle \bullet \rangle = \frac{\int \bullet\, dA}{ \int dA}$ denotes domain-averaging \cite{thuburn2014cascades}. Kinetic energy $(E)$ and enstrophy $(Z)$ are key quantities of interest as they are conserved invariants for unforced, inviscid and drag-free flow \cite{boffetta2012two}.  

\subsection{Resolved kinetic energy, enstrophy, and inter-scale transfers}

For the 2D QG model \eqref{eq:QG}, 
we derive the evolution equations for energy and enstrophy (Appendix\,\ref{sec:AppB}),
\begin{equation}\label{eq:2D_budget}
     \frac{d E}{dt}  =  - \left\langle\psi \frac{d \omega}{dt} \right\rangle, \qquad
     \frac{d Z}{dt}  =    \left\langle\omega \frac{d \omega}{dt}  \right\rangle.
\end{equation}
Using the flux formulation, the resolved energy and enstrophy satisfy the balance equations:
\begin{eqnarray}\label{eq:SGS_budget}
\begin{aligned}
      \frac{d\overline{E}}{dt} -  \langle \fpsi \, \nabla \cdot (\fvel\, \fomega - \bm{\sigma}) \rangle & =  \varepsilon_\nu + \varepsilon_\mu + \varepsilon_F,\\
      \frac{d\overline{Z}}{dt} + \langle \fomega \, \nabla \cdot (\fvel\, \fomega - \bm{\sigma}) \rangle & = \eta_\nu + \eta_\mu + \eta_F.
\end{aligned}
\end{eqnarray}

The right-hand-side terms of
\eqref{eq:SGS_budget} arise from linear interactions and forcing, with
\begin{eqnarray}\label{eq:SGS_budget_2}
\begin{aligned}
\varepsilon_{\nu} = -2\nu\overline{Z},\qquad
\varepsilon_{\mu} = -2\mu\overline{E},\qquad
\varepsilon_{F} = -\langle \fpsi\,\overline{ F}\rangle, \\
\eta_{\nu} = -2\nu\overline{P},\qquad
\eta_{\mu} = -2\mu\overline{Z},\qquad
\eta_{F} = \langle \fomega\,\overline{ F}\rangle,
\end{aligned}
\end{eqnarray}
where $\overline{P} = \frac{1}{2} \langle |\nabla{\fomega}|^2 \rangle$ is the resolved palinstrophy \citep{boffetta2012two, valadao2025spectrum}. Under periodic boundary conditions, the $\beta$-term vanishes after domain averaging and does not directly contribute to the balance equations. To characterize the local effect of the SGS term on the resolved kinetic energy, we decompose the nonlinear contribution using incompressibility as
\begin{equation} 
    \fpsi \, \nabla \cdot (\fvel\, \fomega - \bm{\sigma}) = \nabla \cdot (\fvel \fpsi \fomega - \fpsi \bm{\sigma}) + \bm{\sigma} \cdot \nabla \fpsi.
\end{equation}
The divergence term represents spatial transport (redistribution) of the resolved kinetic energy and vanishes globally, while the second term, 
\begin{equation} \label{eq:SGS_NL_transfer}
    \varepsilon^\ell_{\text{sgs}} = \bm{\sigma} \cdot \nabla \fpsi,
\end{equation}
is the local SGS contribution in the flux formulation. Its spatial average gives the net SGS exchange of resolved kinetic energy, $\varepsilon_{\text{sgs}} = \langle \varepsilon^\ell_{\text{sgs}} \rangle$, while its pointwise value provides a diagnostic of local source and sink regions. This decomposition is analogous to the classical energy transfer term $\tau_{ij}\overline{S}_{ij}$ in velocity–strain formulations, but differs by a divergence term that vanishes globally \cite{eyink2005locality}. The corresponding local SGS enstrophy transfer \cite{eyink2001dissipation, chen2003physical} is $\eta^\ell_{\text{sgs}} = -\bm{\sigma} \cdot \nabla \fomega$. Hence, the subgrid-scale contributions to the resolved energy and enstrophy balance are $\varepsilon_{\text{sgs}} = \langle \varepsilon^\ell_{\text{sgs}} \rangle$ and $\eta_{\text{sgs}} = \langle \eta^\ell_{\text{sgs}} \rangle$.

\section{\label{sec:classical_sgs} Classical Subgrid-scale Models}

We now provide a brief review of classical SGS modeling approaches.

\subsection{Functional Subgrid-Viscosity Models}
Subgrid-viscosity-based models are functional approaches that parameterize $\Pi$ based on the eddy viscosity or Boussinesq hypothesis \cite{sagaut2006large}. Under this hypothesis, the energy or enstrophy transfer between resolved and unresolved scales is assumed 
analogous to 
molecular diffusion. In the vorticity-streamfunction form, we have
\begin{equation} \label{eq:SV_SGS}
    c_{sv}\,\Pi^\text{SV}(\fomega,\fpsi; \Delta_F) = c_{sv}\,\nabla\cdot(\nu_\text{sgs}\nabla\fomega),
\end{equation}
where $\nu_\text{sgs}$ is the eddy viscosity which needs to be modeled, $c_{sv}$ is the model coefficient, and $\Delta_\text{F}$ is the filter width,  which commonly differs from the LES grid resolution, $\Delta_\text{LES}$ (see Appendix\;\ref{sec:AppA}).
We use a conservative form of $\Pi^\text{SV}$ with $\bm{\sigma} = \nu_\text{sgs}\nabla\fomega$, which differs from a purely diffusive form, $\Pi^\text{SV} = \nu_\text{sgs}\nabla^2\fomega$ by the additional term $\nabla\nu_\text{sgs}\cdot\nabla\fomega$. This ensures physical consistency when $\nu_\text{sgs}$ is spatially varying \cite{mansfield1998dynamic}.

One of the most popular subgrid-viscosity approaches is the Smagorinsky model, originally developed from the forward cascade of energy in 3D turbulence \cite{smagorinsky1963general}. Hence, this model assumes an alignment between $\nu_\text{sgs}$ and the filtered strain-rate, $|\overline{S}|$. 
\begin{eqnarray} \label{eq:SV_Smag}
        \nu_\text{sgs} & =  |\overline{S}|, \qquad
        |\overline{S}| & = \sqrt{4 \fpsi_{xy}^2 + (\fpsi_{xx}- \fpsi_{yy})^2} , \qquad  c_{sv}= (C_s\Delta_\text{F})^2.
\end{eqnarray}
The Leith model \cite{leith1996stochastic} is an alternate approach developed based on the forward cascade of enstrophy in 2D turbulence and instead assumes an alignment between $\nu_\text{sgs}$ and the filtered vorticity gradient, $|\nabla{\fomega}|$,
\begin{eqnarray} \label{eq:SV_Leith}
    \nu_\text{sgs} & = |\nabla{\fomega}|, \qquad 
    |\nabla{\fomega}| & = \sqrt{\fomega_x^2 + \fomega_y^2}, \qquad  c_{sv}=  (C_l\Delta_\text{F})^3.
\end{eqnarray}
Although $C_s$ was analytically determined as $0.17$ in 3D isotropic turbulence \cite{lilly1966representation}, $C_s$ and $C_l$ are flow-dependent and are chosen empirically or semi-analytically for 2D flows \cite{canuto1997determination, guan2025semi}. Alternatively, their values can be determined using the dynamic procedure (denoted as the DSmag or DLeith models) based on minimization of the Germano identity error (GIE) \cite{germano1991dynamic, lilly1992proposed}. In general, functional models show low correlation with the \emph{ideal} $\Pi$ in \emph{a priori} tests, and are predominantly dissipative \cite{meneveau2000scale}. In the rest of this work, we use the Leith model for the main functional model. 

\subsection{Structural Nonlinear Gradient Models (NGM)}

Structural models can be obtained either by approximate or iterative deconvolution of the LES filter \cite{stolz1999approximate} or by scale similarity hypotheses \cite{bardina1980improved}. In this work, we focus on the nonlinear gradient model, an approximate deconvolution-based approach obtained as a partial recovery of the SGS forcing using a Taylor series expansion. Deconvolution-based approaches are typically dynamics-agnostic and only applicable to filters with finite moments, such as the box and Gaussian filters. When only the leading-order term in the expansion is retained, the second-order nonlinear gradient model (NGM2) is obtained \cite{leonard1975energy, clark1979evaluation}. For a Gaussian filter of width $\Delta_\text{F}$, it can be expressed in vorticity-streamfunction form as
\begin{equation} \label{eq:NGM2}
    \Pi^\text{NGM2}(\fomega,\fpsi; \Delta_\text{F}) = \frac{\Delta_F^2}{12}  [\overline{\psi}_{xy}\overline{\omega}_{xx} + \overline{\psi}_{yy}\overline{\omega}_{xy} -\overline{\psi}_{xx}\overline{\omega}_{xy} - \overline{\psi}_{xy}\overline{\omega}_{yy}],
\end{equation}
or equivalently, in the flux formulation,
\begin{equation} \label{eq:NGM2_flux}
    \Pi^\text{NGM2}(\fomega,\fpsi; \Delta_\text{F}) = \nabla \cdot \bm{\sigma}, \qquad \bm{\sigma} = - \frac{\Delta_F^2}{12}  [(\nabla \fvel) \,\nabla \fomega ],
\end{equation}
where $\nabla \fvel_{ij} = \partial_j \fvel_i$. From \eqref{eq:NGM2_flux}, the velocity gradient acts as a tensor eddy viscosity in the nonlinear model. This model shows a high correlation with the \emph{ideal} $\Pi$ in \emph{a priori} tests but is known to be deficient because its contribution to the resolved kinetic energy $\varepsilon_{\text{sgs}}$ vanishes in 2D \cite{eyink2006multi,jakhar2024learning}. Therefore, we follow \cite{jakhar2026analytical} and consider the next higher-order correction in the expansion, yielding the fourth-order nonlinear gradient model (NGM4) in vorticity form,
\begin{equation} \label{eq:TD}
    \Pi^\text{NGM4}(\fomega,\fpsi; \Delta_\text{F}) =  \Pi^\text{NGM2} + \frac{\Delta_F^4}{288}  [\overline{\omega}_{xxx}\overline{\psi}_{xxy} - \overline{\omega}_{xxy}(\overline{\psi}_{xxx} - 2\overline{\psi}_{xyy}) + \overline{\omega}_{xyy}(\overline{\psi}_{yyy} - 2\overline{\psi}_{xxy}) - \overline{\omega}_{yyy}\overline{\psi}_{xyy}],
\end{equation}
or equivalently, 
\begin{equation} \label{eq:TD_flux}
    \Pi^\text{NGM4}(\fomega,\fpsi; \Delta_\text{F}) = \nabla \cdot \bm{\sigma}, \qquad \bm{\sigma} = - \frac{\Delta_F^2}{12}  [(\nabla \fvel) \,\nabla \fomega ] - \frac{\Delta_F^4}{288}[
 \nabla \nabla \fvel : \nabla \nabla \fomega],
\end{equation}
where the colon indicates contraction over the two derivative indices, $[
 \nabla \nabla \fvel : \nabla \nabla \fomega]_i
=
\partial_{jk}\fvel_i\,\partial_{jk}\fomega$
\citep{bird2006transport}. 
In the following sections, we use the fourth-order model, NGM4, and refer to it as NGM for brevity. In general, structural models are known to exhibit numerical instabilities due to insufficient dissipation, which can be alleviated by additional regularization or clipping to reduce backscatter \cite{liu1994properties, prakash2022optimal}.

\section{\label{sec:methods} Modified Dynamic Mixed Modeling Framework}

In this section, we show that the standard two-parameter dynamic mixed model leads to structural dominance, and propose a modified framework to develop a new family of dynamic mixed models with a tunable balance between structural and functional contributions. 
Mixed models combine the strengths of structural and functional subgrid-viscosity models and can alleviate the drawbacks of their individual counterparts \cite{bardina1983improved}. For example, inclusion of an eddy viscosity component can regularize and numerically stabilize the structural model \cite{kamal2024artificial}. A general 2-parameter mixed model can be obtained through a linear combination as 
\begin{equation}
    \label{eq:Pi-M}
    \Pi^\text{M}(\fomega,\fpsi; \Delta_\text{F}) = c_{sv} \Pi^\text{SV}(\fomega,\fpsi; \Delta_\text{F}) + c_{ngm} \Pi^\text{NGM}(\fomega,\fpsi; \Delta_\text{F}).
\end{equation}
To avoid ad-hoc combinations, we formulate mixed models through a dynamic procedure based on the application of two filters, a grid filter and a test filter, to determine parameters $c_{sv}$ and $c_{ngm}$ \cite{zang1993dynamic, salvetti1995priori}.

\subsection{Fully-Coupled Dynamic Mixed Model (DMM)}
\label{sec:fcmm}
We first derive the fully-coupled dynamic mixed model \cite{salvetti1995priori, horiuti1997new} following the dynamic procedure in the vorticity-streamfunction form \cite{germano1991dynamic, lilly1992proposed, san2014dynamic}. The term \emph{fully-coupled} emphasizes that no restrictions are imposed on the interactions between the functional and structural components.
We apply a test filter $\widehat{(\bullet)}$ of width $\widehat{\Delta}_\text{F} = 2\,\Delta_\text{F}$ (Appendix\;\ref{sec:AppA}) to \eqref{eq:QG_LES} with $\Pi = \Pi^\text{M}$ \eqref{eq:Pi-M}
resulting in (\ref{eq:QG_LES_Test_1}),
\begin{equation} \label{eq:QG_LES_Test_1}
    \frac{\partial \widehat{\fomega}}{\partial t} + \widehat{J(\fpsi,\fomega)} = \nu \nabla^2 \widehat{\fomega} - \mu \widehat{\fomega} - \beta \widehat{\fpsi}_x + \widehat{\overline{F}} + \widehat{\Pi^\text{M}(\fomega,\fpsi; \Delta_\text{F})}.
\end{equation}
Next, 
we apply the same test filter to \eqref{eq:QG}, and assume $\widehat{\psi} \simeq \widehat{\fpsi}$, $\widehat{\omega} \simeq \widehat{\fomega}$, yielding the corresponding doubly-filtered equation (\ref{eq:QG_LES_Test_2}),
\begin{equation} \label{eq:QG_LES_Test_2}
    \frac{\partial \widehat{\fomega}}{\partial t} + J(\widehat{\fpsi},\widehat{\fomega}) = \nu \nabla^2 \widehat{\fomega} - \mu \widehat{\fomega} - \beta \widehat{\fpsi}_x + \widehat{\overline{F}} + \Pi^\text{M}(\widehat{\fomega},\widehat{\fpsi}; \widehat{\Delta}_\text{F} ).
\end{equation}
Subtracting \eqref{eq:QG_LES_Test_1} from \eqref{eq:QG_LES_Test_2} gives an algebraic relation to be satisfied \eqref{eq:GIE_1}, 
\begin{equation} \label{eq:GIE_1}
    \mathcal{L} = c_{ngm}\mathcal{H} + c_{sv}\mathcal{M},
\end{equation}
where
\begin{eqnarray} \label{eq:LHM_dynamic}
\begin{aligned}
    \mathcal{L} & =  J(\widehat{\fpsi},\widehat{\fomega}) - \widehat{J(\fpsi,\fomega)},\\
    \mathcal{H} & = \Pi^\text{NGM}(\widehat{\fomega},\widehat{\fpsi}; \widehat{\Delta}_F) -\widehat{\Pi^\text{NGM}(\fomega,\fpsi; \Delta_F)},\\
     \mathcal{M} & =  \Pi^\text{SV}(\widehat{\fomega},\widehat{\fpsi}; \widehat{\Delta}_F) - \widehat{\Pi^\text{SV}(\fomega,\fpsi; \Delta_F)}.
\end{aligned}
\end{eqnarray} 
The resulting form of $\mathcal{L}$ \eqref{eq:LHM_dynamic} is analogous to an improved version of the scale similarity model \cite{bardina1980improved} developed in \cite{liu1994properties, sarghini1999scale}. We note that the distinction between the 2D vorticity form of the GIE \eqref{eq:GIE_1} and the standard velocity form of the GIE as introduced in \cite{germano1991dynamic}
affects the well-posedness of the dynamic coefficient estimation problem. In the standard velocity form, the GIE is a tensorial relation and contains multiple components (e.g., three components in 2D), leading to an overdetermined system \cite{germano1991dynamic, lilly1992proposed}. In contrast, the present vorticity formulation reduces the GIE to a single scalar relation \eqref{eq:GIE_1}. Thus, if $c_{ngm}$ and $c_{sv}$ are allowed to vary independently in space, \eqref{eq:GIE_1} provides only one local equation for two unknown coefficients at each spatial location and is therefore underdetermined. If we instead assume that $c_{ngm}$ and $c_{sv}$ are homogeneous in space, with spatially varying eddy viscosity \eqref{eq:SV_SGS} and tensor diffusivity \eqref{eq:TD_flux} by construction, the problem is again overdetermined. 
We can then minimize the domain-averaged squared GIE $\langle E^2 \rangle$ 
with respect to $c_{sv}$ and $c_{ngm}$ in least-squares sense \cite{lilly1992proposed} where 
\begin{equation} \label{eq:GIE_2}
    E = \mathcal{L} - c_{ngm} \mathcal{H} - c_{sv} \mathcal{M}\, .
\end{equation}
Hence, setting ${\partial \langle E^2 \rangle}/{\partial c_{ngm}} = 0$ and ${\partial \langle E^2 \rangle}/{\partial c_{sv}} = 0$ results in a 2-by-2 domain-averaged linear system for $c_{ngm}$ and $c_{sv}$  (\ref{eq:2-DMM}), 
\begin{equation}  \label{eq:2-DMM}
  \underbrace{ \left\langle
        \begin{bmatrix}
        \mathcal{H} & 0  \\
        0 & \mathcal{M}
    \end{bmatrix}^\top
    \begin{bmatrix}
        1 & 1  \\
        1 & 1 
    \end{bmatrix}
    \begin{bmatrix}
        \mathcal{H} & 0  \\
        0 & \mathcal{M}
    \end{bmatrix}
    \right\rangle}_{\mathbf{G} = \langle\mathbf{A}^\top \mathbf{J} \mathbf{A}\rangle}
\,
    \begin{bmatrix}
  c_{ngm} \\ c_{sv}      
    \end{bmatrix}
    = 
   \underbrace{\left\langle
    \begin{bmatrix}
        \mathcal{H} & 0  \\
        0 & \mathcal{M}
    \end{bmatrix}^\top
    \begin{bmatrix}
        \mathcal{L}  \\
        \mathcal{L}
    \end{bmatrix}
      \right\rangle}_{\langle\mathbf{A}^\top \mathbf{L}\rangle},
\end{equation}
where $\mathbf{J}$ is the standard all-ones matrix,
$\mathbf{G} = \langle\mathbf{A}^\top \mathbf{J} \mathbf{A}\rangle$ is the Gram matrix formed by the domain-averaged inner products of the functional and structural contributions to the GIE,
and $\langle \mathbf{A}^\top \mathbf{L}\rangle$ is the projection of the NGM and SV components onto $\mathcal{L}$.
All domain-averaged variables in \eqref{eq:2-DMM} can be positively clipped with a Heaviside function to remove negative $c_{sv}$ values, thereby preventing excessive backscatter and numerical instabilities \cite{meneveau2000scale}. 

\subsection{Family of Modified Dynamic Mixed Models}
The least-squares solution of the fully-coupled system \eqref{eq:2-DMM} is typically biased toward the structural term {$\Pi^\text{NGM}$}, leaving the functional contribution {$\Pi^\text{SV}$} too weak to provide effective dissipative regularization \cite{anderson1999effects}. 
This structural dominance can be understood from a least-squares perspective by analyzing the projections of each component's contribution to the GIE onto $\mathcal{L}$,
i.e., the right hand side of \eqref{eq:2-DMM}. To illustrate, we consider the limiting case where the improved scale-similarity model ($\Pi^\text{SIM}$) \cite{liu1994properties,sarghini1999scale} is used as the structural component in \eqref{eq:Pi-M}, with dynamic coefficient $c_{sim}$. Since $\mathcal{L}$ is analogous to the improved scale-similarity model evaluated at the test-filter level, we have $\Pi^\text{SIM} \to \mathcal{L}$. Assuming the test filter also lies within the inertial range over which scale-similarity arguments remain valid, 
the contribution of $\Pi^\text{SIM}$ to the GIE, which we denote as $\mathcal{H}_\text{SIM}$, is expected to have the same leading-order structure as $\mathcal{L}$ 
such that $\mathcal{H}_\text{SIM} \approx \gamma \mathcal{L}$, where $\gamma \sim \mathcal{O}(1)$ accounts for differences in amplitude introduced by the two filter levels \cite{cook1997determination, meneveau2000scale}. Substituting $\Pi^\text{SIM} \to \mathcal{L}$ and $\mathcal{H}_\text{SIM} \approx \gamma \mathcal{L}$ into the domain-averaged least-squares system
\eqref{eq:2-DMM}, it reduces to
\begin{equation}
   \begin{bmatrix}
        \gamma^2 \langle\mathcal{L}^2\rangle & \gamma \langle\mathcal{L}\mathcal{M}\rangle \\
        \gamma \langle\mathcal{L}\mathcal{M}\rangle & \langle\mathcal{M}^2\rangle 
    \end{bmatrix}
\,
    \begin{bmatrix}
  c_{sim} \\ c_{sv}      
    \end{bmatrix}
    = 
    \begin{bmatrix}
       \gamma  \langle\mathcal{L}^2\rangle  \\
        \langle\mathcal{L}\mathcal{M}\rangle 
    \end{bmatrix},
    \label{eq:2-DMM_Limit}
\end{equation}
and most of the fit is assigned to the structural term (when non-singular), i.e., the solution is $c_{\mathrm{sim}} \approx 1/\gamma$ with a subgrid-viscosity contribution $c_{\mathrm{sv}} \approx 0$. Thus, in this limiting case, the least-squares procedure assigns no contribution to the subgrid-viscosity term. Finally, since the nonlinear gradient model {$\Pi^\text{NGM}$} is a higher-order Taylor-series approximation to the scale-similarity model, $\Pi^\text{SIM}$ \cite{pruett2001taylor}, its GIE contribution $\mathcal{H}$ \eqref{eq:LHM_dynamic} is similarly expected to project more strongly onto $\mathcal{L}$ than $\mathcal{M}$. Therefore, even though the mixed model contains both components, the dynamically obtained closure can behave as a structurally dominated model rather than a balanced mixed model. This motivates modifications to the dynamic procedure to enable the control of the coupling between the structural and functional components. 

We propose such modifications by decomposing the Gram matrix ${\mathbf{G}}$ in \eqref{eq:2-DMM} based on the interactions between the subgrid-viscosity and nonlinear gradient terms: their self-interactions are the diagonal entries while their cross-interactions are the off-diagonal entries,
\begin{equation}
   \underbrace{\begin{bmatrix}
        \langle\mathcal{H}^2\rangle & \langle\mathcal{H}\mathcal{M}\rangle \\
        \langle\mathcal{H}\mathcal{M}\rangle & \langle\mathcal{M}^2\rangle 
    \end{bmatrix}}_{\mathbf{G}}
\,
    =
     \underbrace{
    \begin{bmatrix}
        1 & 0 \\
        0 & 0 
    \end{bmatrix}
    \langle\mathcal{H}^2\rangle}
    _{\text{Nonlinear Gradient (\ref{eq:TD})}} 
    +
    \underbrace{
    \begin{bmatrix}
        0 & 0 \\
        0 & 1 
    \end{bmatrix}
    \langle\mathcal{M}^2\rangle}
    _{\text{Subgrid-Viscosity (\ref{eq:SV_SGS})}}
    +
    \underbrace{
    \begin{bmatrix}
        0 & 1\\
        1 & 0
    \end{bmatrix}
    \langle\mathcal{H}\mathcal{M}\rangle}
    _{\text{Cross-interactions}}.
    \label{eq:2-DMM_reduction}
\end{equation}
The off-diagonal entries determine how the structural and functional components compete in the dynamic procedure. When the structural component more strongly projects onto $\mathcal{L}$, the fully-coupled system can assign most of the contribution to the structural term, thereby reducing the regularizing influence of the 
subgrid-viscosity.

Thus, we construct a novel family of modified dynamic mixed models by selectively restricting specific cross-interactions in \eqref{eq:2-DMM_reduction}. This is achieved by replacing the standard all-ones matrix $\mathbf{J}$ in \eqref{eq:2-DMM} with a coupling matrix $\mathbf{\Delta}$ where the off-diagonal elements are composed of modifiers $\delta_{12},\,\delta_{21}$, leading to a modified Gram-based coefficient matrix,
\begin{equation} \label{eq:Gmod_ADeltaA}
    \underbrace{ \left\langle
    \begin{bmatrix}
        \mathcal{H} & 0  \\
        0 & \mathcal{M}
    \end{bmatrix}^\top
    \begin{bmatrix}
        1 & \delta_{12}  \\
        \delta_{21}  & 1 
    \end{bmatrix}
    \begin{bmatrix}
        \mathcal{H} & 0  \\
        0 & \mathcal{M}
    \end{bmatrix}
    \right\rangle }_{\langle\mathbf{A}^\top \mathbf{\Delta} \mathbf{A}\rangle}
\, 
=      
    \underbrace{\begin{bmatrix}
        \langle\mathcal{H}^2\rangle & \delta_{12}\langle\mathcal{H}\mathcal{M}\rangle \\
        \delta_{21}\langle\mathcal{H}\mathcal{M}\rangle & \langle\mathcal{M}^2\rangle 
    \end{bmatrix}}_{\mathbf{G}_{mod}},
\,
\end{equation}
where
\begin{equation}
    \underbrace{\begin{bmatrix}
        \langle\mathcal{H}^2\rangle & \delta_{12}\langle\mathcal{H}\mathcal{M}\rangle \\
        \delta_{21}\langle\mathcal{H}\mathcal{M}\rangle & \langle\mathcal{M}^2\rangle 
    \end{bmatrix}}_{\mathbf{G}_{mod}}
\,
    =
    \underbrace{
    \begin{bmatrix}
        1 & 0 \\
        0 & 0 
    \end{bmatrix}
    \langle\mathcal{H}^2\rangle}
    _{\text{Nonlinear Gradient (\ref{eq:TD})}} 
    +
    \underbrace{
    \begin{bmatrix}
        0 & 0 \\
        0 & 1 
    \end{bmatrix}
    \langle\mathcal{M}^2\rangle}
    _{\text{Subgrid-Viscosity (\ref{eq:SV_SGS})}}
    +
    \underbrace{
    \begin{bmatrix}
        0 & \delta_{12}\\
        \delta_{21} & 0
    \end{bmatrix}
    \langle\mathcal{H}\mathcal{M}\rangle}
    _{\text{Modified cross-interactions}}.
    \label{eq:2-DMM_mod}
\end{equation}
Without loss of generality,
the values of
$\delta_{12}$ and $\delta_{21}$
can be limited to be between 0 and 1.
The coupling matrix $\mathbf{\Delta}$ governs the least-squares projection of $\mathcal{L}$ onto the span of $\{\mathcal{H},\mathcal{M}\}$. Its symmetric component, $\mathbf{\Delta}_{sym} = (\mathbf{\Delta}^\top + \mathbf{\Delta})/2$, acts as a metric whose off-diagonal elements, $(\delta_{12} + \delta_{21})/2$, define the degree of non-orthogonality between $\mathcal{H}$ and $\mathcal{M}$.  
Its antisymmetric component, $(\delta_{12} - \delta_{21})/2$, introduces directional coupling.
Further, the modified Gram-based coefficient matrix $\mathbf{G}_{mod}$
\eqref{eq:Gmod_ADeltaA} can be interpreted as the result of two coupled least-squares problems, one for each coefficient: ${\partial \langle E_{\mathcal{H}}^2 \rangle}/{\partial c_{ngm}} = 0$ and ${\partial \langle E_{\mathcal{M}}^2 \rangle}/{\partial c_{sv}} = 0$, where,
\begin{eqnarray} \label{eq:modified_GIE_1}
    E_{\mathcal{H}} & = \mathcal{L} - c_{ngm} \mathcal{H} -  \delta_{12} c_{sv} \mathcal{M},\\ \label{eq:modified_GIE_2}
    E_{\mathcal{M}} & = \mathcal{L} - \delta_{21} c_{ngm} \mathcal{H} - c_{sv} \mathcal{M}.
\end{eqnarray}
Hence, $\delta_{12}$ controls how strongly the subgrid-viscosity contribution $\mathcal{M}$ affects the optimization of the structural coefficient $c_{ngm}$, and similarly $\delta_{21}$ controls how strongly the structural contribution $\mathcal{H}$ affects the optimization of the functional coefficient $c_{sv}$. When $\delta_{12} = \delta_{21} = 1$, we recover $E_{\mathcal{H}} = E_{\mathcal{M}} = E$, leading back to \eqref{eq:2-DMM}. Hence, the modified system of equations to be solved is
\begin{equation} \label{eq:Gmod_linsys}
   \underbrace{\begin{bmatrix}
        \langle\mathcal{H}^2\rangle & \delta_{12} \langle\mathcal{H}\mathcal{M}\rangle \\
        \delta_{21} \langle\mathcal{H}\mathcal{M}\rangle & \langle\mathcal{M}^2\rangle 
    \end{bmatrix}}_{\mathbf{G}_{mod}}
\,
    \begin{bmatrix}
  c_{ngm} \\ c_{sv}      
    \end{bmatrix}
    = 
    \begin{bmatrix}
        \langle\mathcal{L}\mathcal{H}\rangle  \\
        \langle\mathcal{L}\mathcal{M}\rangle 
    \end{bmatrix}.
\end{equation}

For the rest of this work, we focus on classes of modified mixed models developed using binary modifiers ($\delta_{12},\,\delta_{21} \in \{0, 1\}$), as described below. 
Results for modifiers that are allowed to vary continuously ($0 < \delta < 1$) are summarized in Appendix\;\ref{sec:AppC}. 

\subsubsection{Sequential Dynamic Mixed Models (SDMM-1, SDMM-2)} \label{sec:sdmm}

The first class 
is the sequential dynamic mixed models. The modifiers are set to selectively limit cross-interactions, leading to two sequential dynamic mixed models denoted as SDMM-1 (when $\delta_{12} = 0, \,\delta_{21} = 1$) and SDMM-2 (when $\delta_{12} = 1, \,\delta_{21} = 0$). 
In SDMM-1, contributions from the subgrid-viscosity component are prevented from directly influencing the dynamic nonlinear gradient component, yielding
\begin{equation}
   \begin{bmatrix}
        \langle\mathcal{H}^2\rangle & 0\\
        \langle\mathcal{H}\mathcal{M}\rangle & \langle\mathcal{M}^2\rangle 
    \end{bmatrix}
\,
    \begin{bmatrix}
  c_{ngm} \\ c_{sv}      
    \end{bmatrix}
    = 
    \begin{bmatrix}
        \langle\mathcal{L}\mathcal{H}\rangle  \\
        \langle\mathcal{L}\mathcal{M}\rangle 
    \end{bmatrix}.
    \label{eq:SDMM-1}
\end{equation}
Therefore, $c_{ngm}$ is obtained dynamically first (${\partial \langle E_{\mathcal{H}}^2 \rangle}/{\partial c_{ngm}} = 0$), which is then corrected by the subgrid-viscosity model (${\partial \langle E_{\mathcal{M}}^2 \rangle}/{\partial c_{sv}}={\partial \langle E^2 \rangle}/{\partial c_{sv}} = 0$), resulting in much smaller values of $c_{sv}$ than the DSmag or DLeith procedures. Accordingly, the structural component remains dominant as in \eqref{eq:2-DMM}. On the other hand, in SDMM-2, the values of $c_{sv}$ are first obtained identically to the DSmag or DLeith procedures (${\partial \langle E_{\mathcal{M}}^2 \rangle}/{\partial c_{sv}} = 0$), followed with a correction by the dynamic nonlinear gradient model (${\partial \langle E_{\mathcal{H}}^2 \rangle}/{\partial c_{ngm}} = {\partial \langle E^2 \rangle}/{\partial c_{ngm}} = 0$),
\begin{equation}
   \begin{bmatrix}
        \langle\mathcal{H}^2\rangle & \langle\mathcal{H}\mathcal{M}\rangle \\
        0 & \langle\mathcal{M}^2\rangle 
    \end{bmatrix}
\,
    \begin{bmatrix}
  c_{ngm} \\ c_{sv}      
    \end{bmatrix}
    = 
    \begin{bmatrix}
        \langle\mathcal{L}\mathcal{H}\rangle  \\
        \langle\mathcal{L}\mathcal{M}\rangle 
    \end{bmatrix}.
    \label{eq:SDMM-2}
\end{equation}
Hence, SDMM-2 leads to a smaller value of $c_{ngm}$ compared to SDMM-1, with the functional model playing a stronger role. While a similar model was obtained in \cite{morinishi2001recommended} for wall-bounded flows through an approximation based on the condition number of ${\mathbf{G}}$, here the modified model arises naturally from the Gram-matrix decomposition \eqref{eq:Gmod_ADeltaA}-\eqref{eq:2-DMM_mod}. 

\subsubsection{Fully-Decoupled Dynamic Mixed Models: FDMM($\alpha$)} \label{sec:fdmm}

Setting both $\delta_{12} = 0$ and $\delta_{21} = 0$ yields a fully-decoupled dynamic procedure where $c_{sv}$ and $c_{ngm}$ are obtained dynamically yet independently (${\partial \langle E_{\mathcal{H}}^2 \rangle}/{\partial c_{ngm}} = 0$ and ${\partial \langle E_{\mathcal{M}}^2 \rangle}/{\partial c_{sv}} = 0$). The result is a final class of modified dynamic mixed models denoted as FDMM($\alpha$),
\begin{subequations}\label{eq:FDMM}
\begin{gather}
\begin{bmatrix}
\langle\mathcal{H}^2\rangle & 0 \\
0 & \langle\mathcal{M}^2\rangle 
\end{bmatrix}
\begin{bmatrix}
c_{ngm} \\ c_{sv}
\end{bmatrix}
 =
\begin{bmatrix}
\langle\mathcal{L}\mathcal{H}\rangle  \\
\langle\mathcal{L}\mathcal{M}\rangle 
\end{bmatrix}, \label{eq:FDMM_coeffs}\\
\Pi^{\text{FDMM}(\alpha)}(\fomega,\fpsi; \Delta_\text{F})
= \alpha c_{sv} \Pi^\text{SV}(\fomega,\fpsi; \Delta_\text{F})
+ (1-\alpha) c_{ngm} \Pi^\text{NGM}(\fomega,\fpsi; \Delta_\text{F}), \label{eq:FDMM_model}
\end{gather}
\end{subequations}
where $\alpha$ is a mixing parameter that can be chosen \emph{a priori} to set the relative weights of the functional and structural contributions, while the dynamic coefficients are still estimated from the instantaneous GIE \eqref{eq:modified_GIE_1}-\eqref{eq:modified_GIE_2}. Hence, FDMM($\alpha$) is a natural parametric extension of the one-parameter dynamic mixed models proposed in \cite{zang1993dynamic, vreman1996large, winckelmans2001explicit}. In contrast to the fully-coupled formulation, where the structural component often dominates, the FDMM($\alpha$) framework enables explicit tuning of the balance between dissipative and structural mechanisms. These models, therefore, provide insight into the relative roles of the functional and structural components. When $\alpha=1$, FDMM($\alpha$) reduces to the purely functional dynamic model (e.g., DSmag or DLeith), while $\alpha=0$ yields the purely structural dynamic nonlinear gradient model as in \cite{winckelmans2001explicit}. Intermediate values of $\alpha$ yield balanced mixed models, in which both components contribute in a controlled and interpretable manner. More generally, any subset of the mixing or model coefficients can be prescribed \emph{a priori}, with the remaining unknown parameters estimated using GIE minimization \cite{sagaut2000general}. This additional flexibility is not exemplified in this work.

In summary, the novel family of modified dynamic mixed models introduced here (Table\;\ref{tab:methods_models}) spans a range of coupling between functional and structural closures. The fully-coupled dynamic mixed model represents the most general formulation, allowing unrestricted interactions between components. The sequential dynamic mixed models selectively restrict cross-interactions, introducing an implicit hierarchy between components. Finally, the fully-decoupled FDMM($\alpha$) formulation eliminates cross-interactions and enables explicit control over the relative contributions of the functional and structural parts. We next describe the numerical methods and test cases used to evaluate these closures.

\begin{table}[]
\caption{\label{tab:methods_models}
Summary of modified dynamic mixed subgrid-scale models and their descriptions.
}
\begin{ruledtabular}
\begin{tabular}{c|cccl}
 Model & $\delta_{12}$ & $\delta_{21}$ & $\alpha$ & Description  \\
\colrule
DLeith & 0 & 0 & 1 & Dynamic Leith Model\\
DNGM & 0 & 0 & 0 & Dynamic $4^{th}$-order Nonlinear Gradient Model \\
DMM & 1 & 1 & $-$ & Fully-Coupled Dynamic Mixed Model\\
SDMM-1 & 0 & 1 & $-$ & \makecell[l]{Sequential Dynamic Mixed Model\\
{\footnotesize (Functional correction to structural component)}}\\
SDMM-2 & 1 & 0 & $-$ & \makecell[l]{Sequential Dynamic Mixed Model\\
{\footnotesize (Structural correction to functional component)}}\\
FDMM($\alpha$) & 0 & 0 & $0 < \alpha < 1$ & \makecell[l]{Fully-Decoupled Dynamic Mixed Model \\
{\footnotesize (Tunable functional and structural contributions)}}
\end{tabular}
\end{ruledtabular}
\end{table}

\section{\label{sec:test_cases} Numerical Methods and Test Cases}

\begin{figure}[!t]
    \centering
    \includegraphics[width=1\linewidth]{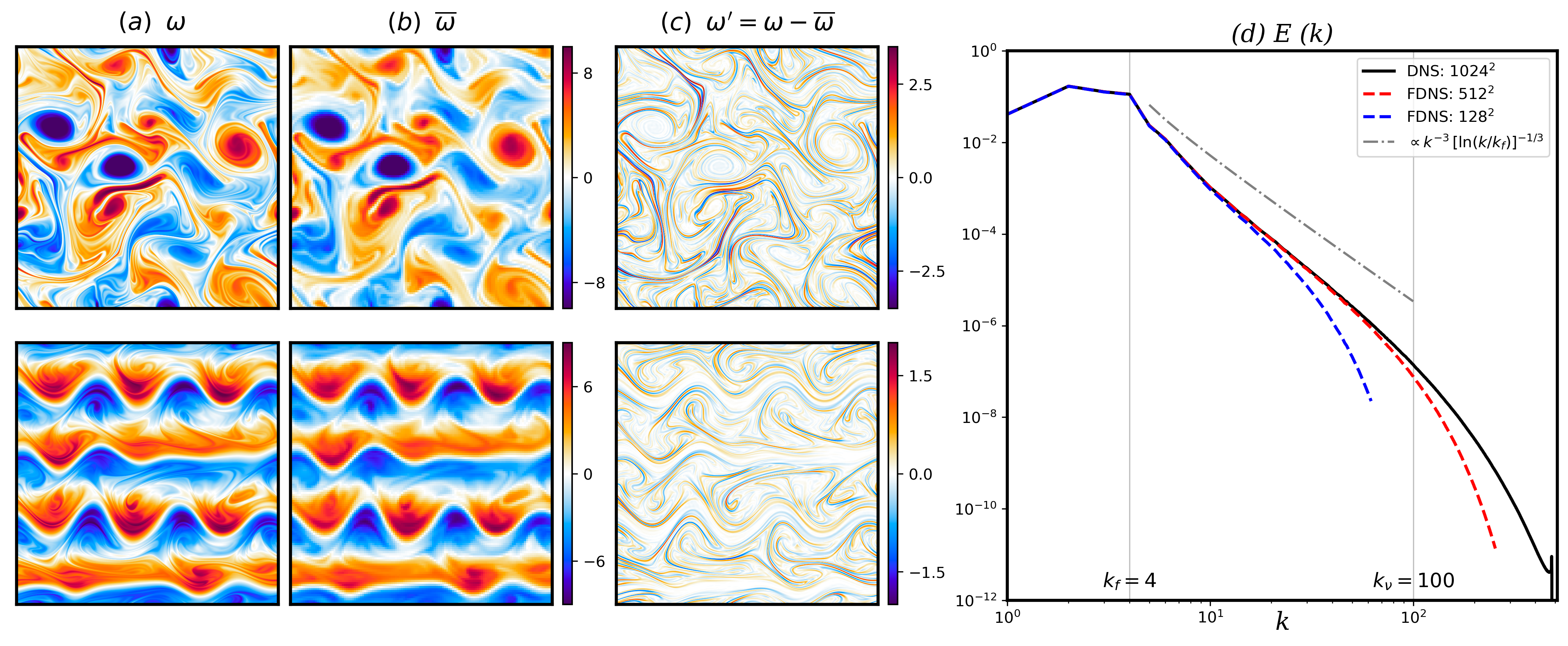}
    \caption{Instantaneous snapshots of (a) DNS vorticity $\omega$ (grid resolution of $1024^2$), (b) filtered vorticity $\fomega$ (resolution of $128^2$), and (c) SGS vorticity $\omega' = \omega - \fomega$. The top row corresponds to the eddy regime ($\beta = 0)$ and the bottom row to the jet regime ($\beta = 10)$. (d) The corresponding angle-averaged kinetic energy spectra $E(k)$ for the eddy regime, comparing DNS  with the filtered DNS (FDNS) fields at resolutions of $512^2$ and $128^2$.}
    \label{fig:test_cases}
\end{figure}

To solve \eqref{eq:QG} and \eqref{eq:QG_LES} numerically, we use the standard Fourier pseudo-spectral spatial discretization on a square, doubly-periodic domain of size $[2 \pi, 2\pi]$. Linear terms are evaluated in Fourier space and nonlinear terms in physical space with 2/3-rule dealiasing \cite{orszag1974numerical}. For a grid of $N^2$ resolution, the maximum resolved numerical cut-off wavenumber is thus $k_{max} =  N/3$ \cite{wan2010accuracy}. For time-stepping, we treat linear terms implicitly with a  Crank-Nicolson scheme and nonlinear terms explicitly with a second-order Adams-Bashforth scheme, leading to an IMEX AB2-CN scheme \cite{boyd2001chebyshev}. The solver is GPU-compatible and developed in PyTorch \cite{suresh_babu_et_al_JAMES2026, suresh_babu_et_al_Oceans2025}.

Following \cite{graham2013framework, frezat2022posteriori}, we non-dimensionalize \eqref{eq:QG} using a length scale $l_0 = 504 \times 10^4/ \pi$\;m and time scale $t_0 = 1.2 \times 10^6$\;s, corresponding to geophysical mesoscales. For the high-fidelity baseline, we use a fine grid with resolution $N_0^2 = 1024^2$, i.e., $(k_{max} \approx 341)$. In non-dimensional form, we use $\mu$ = 0.1 and $\nu = 4 \times 10^{-5}$. We consider two dynamical regimes, the eddy regime with $\beta = 0$ and the jet regime with $\beta = 10$. The flow is driven at a length scale $l_f = 2 \pi/ k_f$ with sinusoidal forcing,
\begin{equation}  \label{eq:forcing}
    F = C(t)[\cos(k_f\;y) - \cos(k_f\;x)],
\end{equation}
where $k_f = 4$, and $C(t)$ is dynamically adjusted to maintain a constant enstrophy injection rate $\langle F \omega  \rangle = \frac{3}{2}$, producing statistically stationary turbulence. 
Vorticity fields are initialized by sampling from a standard Gaussian distribution with wavenumbers $k_{\text{init}}\in [3,10]$ \cite{mcwilliams1999vortices}. A non-dimensional time-step $dt = 5\times 10^{-5}$ is used to simulate up to $T = 100$. We monitor the domain-averaged kinetic energy and enstrophy \eqref{eq:2D_EZP} and observe self-similarity in the spectra at $T \approx 40$, which corresponds to approximately $150$ eddy turnover periods, where the eddy turnover period is defined as
$T_e = 1/\sqrt{2Z}$ \cite{bracco2010reynolds}. We discard the spin-up and obtain the time-averaged root-mean-square velocity, $u_{rms} = \sqrt{2 E} \approx 1$. We define the forcing-scale Reynolds number following \cite{graham2013framework}, $Re_f = \frac{u_{rms}\,l_f}{\nu} \approx 40,000$. Since 2D turbulence shows a forward cascade of enstrophy, we use the time-averaged palinstrophy to define a wavenumber corresponding to the onset of dissipation \cite{bracco2010reynolds, vinograd2026dimensional} as
\begin{equation}\label{eq:dissip_range}
        k_\nu = \left(\frac{2 P}{\nu^2}\right)^{1/6}.
\end{equation}
For both dynamical regimes, we obtain $k_{\nu} \approx 100 < 0.5\,k_{max}$, indicating that the dissipation range is well separated from the effective numerical cut-off $k_{max}$ \cite{xie2020modeling}. We therefore treat the $1024^2$ simulation as the DNS reference. 

Filtered variables are obtained using a Gaussian filter, followed by coarse-graining to the LES grid using a sharp spectral cut-off (Appendix\;\ref{sec:AppA}).
To assess SGS model performance across filtering scales \cite{pope2004ten}, we consider two LES resolutions, $N_{\mathrm{LES}}^2= 128^2$ and $N_{\mathrm{LES}}^2= 512^2$. For the $128^2$ resolution, the cut-off wavenumber is $k_c = 64 < k_{\nu}$, indicating the filter cut-off lies within the inertial range. In contrast, for the $512^2$ resolution, $k_c = 256 > k_{\nu}$, placing the filter cut-off within the dissipative range. However, since the Gaussian filter acts smoothly in spectral space, the filtered fields still exhibit mild attenuation near the upper end of the inertial range before sharp spectral truncation.

Fig.\;\ref{fig:test_cases} shows a comparison of the DNS vorticity field with the filtered vorticity field on a coarse grid of $128^2$ resolution for the eddy and jet regimes. The SGS vorticity field, $\omega' = \omega - \fomega$, highlights the fine-scale structures lost by filtering. In the eddy regime, the SGS vorticity is dominated by thin filamentary structures surrounding coherent vortices, while in the jet regime, they are organized along the zonal jets. The corresponding kinetic energy spectrum is computed as
\begin{equation} \label{eq:E_spectrum}
E(k)
=
\frac{1}{2}
\oiint_{|\mathbf{k}|=k}
k^2\, |\Tilde{\psi}(\mathbf{k})|^2 \, dS,
\end{equation}
where $\oiint (\bullet) dS$ denotes a surface integral over the wavenumber shell of radius $|\mathbf{k}|=k$. From Fig.\;\ref{fig:test_cases} (d), the $128^2$ filtering removes a substantial portion of energy from the inertial range while the $512^2$ filtering primarily removes energy from the dissipative range with weak removal of the inertial range energy. Due to the presence of friction, the inertial range spectrum of the DNS exhibits a slope slightly steeper than the classical $k^{-3}$ power-law \cite{valadao2025spectrum}, but remains consistent with Kraichnan's logarithmic correction $ln(k/k_f)^{-1/3}$ \cite{kraichnan1971inertial}.

\section{\label{sec:results} Results and Discussion}

We perform both \emph{a priori} and \emph{a posteriori} tests to evaluate the performance of the SGS models. For all tests shown, we utilize the Leith model \eqref{eq:SV_Leith} as the functional component and the fourth-order nonlinear gradient model \eqref{eq:TD}-\eqref{eq:TD_flux} as the structural component.
We note that when we used the Smagorinsky model as the functional component, we obtained similar results (not shown), indicating that our main findings are not specific to the subgrid-viscosity closure. 
No clipping is applied to the models or the dynamically estimated coefficients. In all cases, the coefficients $c_{ngm}$ and $c_{sv}$ are obtained by solving \eqref{eq:Gmod_linsys} 
given the
values of $\delta_{12}$ and $\delta_{21}$ (as listed in Table \ref{tab:methods_models}). The coefficients are then substituted into the closure form  \eqref{eq:Pi-M}, except for FDMM($\alpha$) models, which use the modified form \eqref{eq:FDMM_model}.
Since many tests for SGS model performance are statistical, we use the Pattern Correlation Coefficient (CC) \cite{salvetti1995priori, lermusiaux_MWR1999} between a modeled field $\phi^{M}$ and its ideal counterpart $\phi^{FDNS}$, given by 
\begin{equation} \label{eq:PCC}
    \text{CC}(\phi^{FDNS},\phi^M) = \frac{\average{(\phi^M - \average{\phi^M }) (\phi^{FDNS} - \average{\phi^{FDNS}})}}{\average{(\phi^M  - \average{\phi^M })^2}^{1/2} \average{(\phi^{FDNS} - \average{\phi^{FDNS}})^2}^{1/2}} .
\end{equation}

\subsection{A priori tests}
In \emph{a priori} tests, filtered DNS (FDNS) data is utilized to evaluate the modeled SGS forcing ($\Pi^M$), which is then compared against the ideal SGS forcing ($\Pi^{FDNS}$). Hence, these tests do not integrate the LES equations. As part of the \emph{a priori} tests, we examine the temporal evolution of the dynamic coefficients, the structural fidelity of the SGS forcing, and the local and statistical properties of the SGS energy exchange, $\varepsilon^\ell_{\text{sgs}}$.

\subsubsection{Evolution  of dynamic coefficients}
\begin{figure}[!t]
    \centering
    \includegraphics[width=0.9\linewidth]{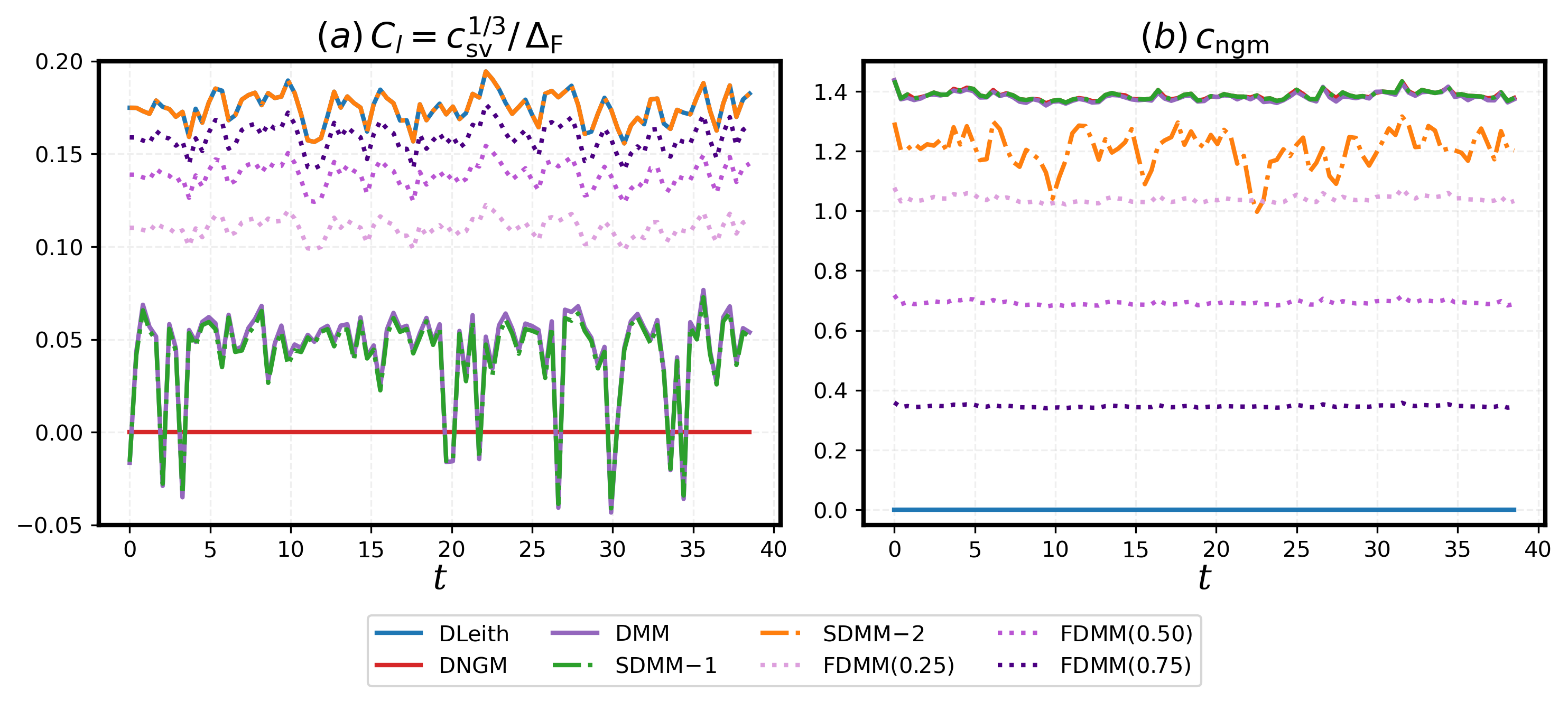}
    \caption{
    \emph{A priori} tests in the eddy regime (resolution of $128^2$):
    Evolution of the dynamic coefficients (a) $C_{l}=c_{sv}^{1/3}/\Delta_{F}$ and (b) $c_{ngm}$ associated with the functional Leith and structural fourth-order nonlinear gradient components, respectively.}
    \label{fig:aprio_coeff}
\end{figure}

We study the temporal evolution of the estimated dynamic coefficients and assess how the modified dynamic procedures alter the relative weights of the functional and structural components. Fig.\;\ref{fig:aprio_coeff} shows the evolution for the eddy regime at a resolution of $128^2$. For the functional component, DLeith and the sequential dynamic mixed model SDMM-2 yield identical values of $c_{sv}$ in \emph{a priori} tests as expected from the construction of the sequential procedure \eqref{eq:SDMM-2}. 
These values are close to those obtained 
in \cite{guan2025semi}. In contrast, the structural component dominates the mixed model in the fully-coupled DMM and sequential dynamic mixed model SDMM-1. In some instances, this leads to slightly negative values of $c_{sv}$ with very small magnitude. For the structural component, identical dynamic values of $c_{ngm}$ are obtained for DNGM and SDMM-1, again as expected from \eqref{eq:SDMM-1}. 
The fully-coupled DMM values are nearly indistinguishable in this case because the solution is structurally dominated and its dynamically obtained $c_{sv}$ values are very small. The corresponding values of $c_{ngm}$ remain larger than unity, in agreement with previous findings \cite{anderson1999effects, winckelmans2001explicit}. 
Here, SDMM-2 slightly shifts the mixed model toward the functional component, resulting in reduced $c_{ngm}$ values closer to unity. Similar behavior is observed in the jet regime (not shown).
The coefficients of the fully-decoupled FDMM($\alpha$) models follow the same trends as their corresponding single-component dynamic models (DLeith and DNGM) but are scaled by the mixing parameter $\alpha$. This direct scaling occurs in \emph{a priori} tests since all closures are evaluated on the same filtered fields. At finer LES resolutions, the values of $c_{sv}$ for all models increase moderately, while those of $c_{ngm}$ decrease in both regimes (not shown). 
This behavior is consistent with the filter cut-off ($k_c$) moving into the dissipative range ($k_\nu$) \eqref{eq:dissip_range} as the resolution is refined (Fig.\;\ref{fig:test_cases}). As $k_c \to k_\nu$, inertial-range scale-similarity arguments are less applicable \cite{meneveau1997dynamic, pope2004ten}, weakening the relative contribution of the structural term and increasing the subgrid-viscosity coefficient. We also observe that the net magnitude of the SGS forcing decreases at finer resolutions. Overall, our modified dynamic framework spans a broad range of functional and structural coefficient combinations, allowing the roles of structural and functional components in a mixed model to be systematically isolated and assessed.

\subsubsection{Structural fidelity of $\Pi^{M}$}

\begin{table}[h]
\caption{\label{tab:aprio_sgs}
\emph{A priori} comparison of correlations of the SGS forcing, $\text{CC}\;(\Pi^{FDNS},\Pi^M)$
}
\begin{ruledtabular}
\begin{tabular}{c|cccccccc}
 & DLeith & DNGM & DMM & SDMM-1 & SDMM-2 & FDMM(0.25) &  FDMM(0.5) & FDMM(0.75)  \\
\colrule
Eddy, $128^2$ & 0.45 & 0.98 & 0.98 & 0.98 & 0.96 & 0.98 & 0.97 & 0.90 \\
Jet, $128^2$ & 0.27 & 0.99 & 0.99 & 0.99 & 0.99 & 0.99 & 0.99 & 0.98 \\
Eddy, $512^2$ & 0.61 & 0.99 & 0.99 & 0.99 & 0.97 & 0.99 & 0.98 & 0.91 \\
Jet, $512^2$ & 0.60 & 0.99 & 0.99 & 0.99 & 0.98 & 0.99 & 0.98 & 0.93 
\end{tabular}
\end{ruledtabular}
\end{table}

Next, the ability of the SGS models to capture the spatial structure of the ideal SGS term is quantified using the Pattern Correlation Coefficient, $\text{CC}\;(\Pi^{FDNS},\Pi^M)$ \eqref{eq:PCC}. Table\;\ref{tab:aprio_sgs} shows that models with a very small functional coefficient ($c_{sv}$) in Fig.\;\ref{fig:aprio_coeff} (i.e, DNGM, DMM, and SDMM-1) are strongly correlated (CC $\geq 0.98$) with the ideal SGS forcing for all dynamical regimes and resolutions considered. This is expected since experimental and theoretical studies have shown that the gradient model reproduces the local geometric structure and alignment properties of the ideal SGS forcing \cite{tao2002statistical, chen2003physical}. The purely functional DLeith model shows substantially weaker correlation, particularly in the jet regime at $128^2$ resolution. The modified mixed models span the range between these limits. While SDMM-1 shows similar performance to DNGM and DMM, the SDMM-2 and FDMM($\alpha$) models retain high correlations (CC $\geq 0.9$) but show a systematic reduction in structural fidelity as the functional contribution increases.

\subsubsection{Subgrid-scale energy exchange}

We now assess the accuracy of the local SGS energy exchange. 
Backscatter refers to the inter-scale energy transfer by the SGS forcing, from the unresolved subgrid scales to the resolved large scales.
While excessive backscatter can lead to numerical instabilities, its inclusion is often necessary to prevent over-dissipation and maintain realistic energy transfers to the resolved scales \cite{jansen2014parameterizing}. From \eqref{eq:SGS_NL_transfer}, we utilize the local SGS contribution to the resolved kinetic energy, $\varepsilon^\ell_{\text{sgs}}$, as a spatial indicator of backscatter \cite{beaumard2026derivation}. At each spatial location, $\varepsilon^\ell_{\text{sgs}}>0$  indicates that the SGS term locally acts as a source of resolved kinetic energy (backscatter), and $\varepsilon^\ell_{\text{sgs}} < 0$ indicates that it acts as a local sink (forward scatter to subgrid-scales).

\begin{figure}[h!]
    \centering
    \includegraphics[width=1\linewidth]{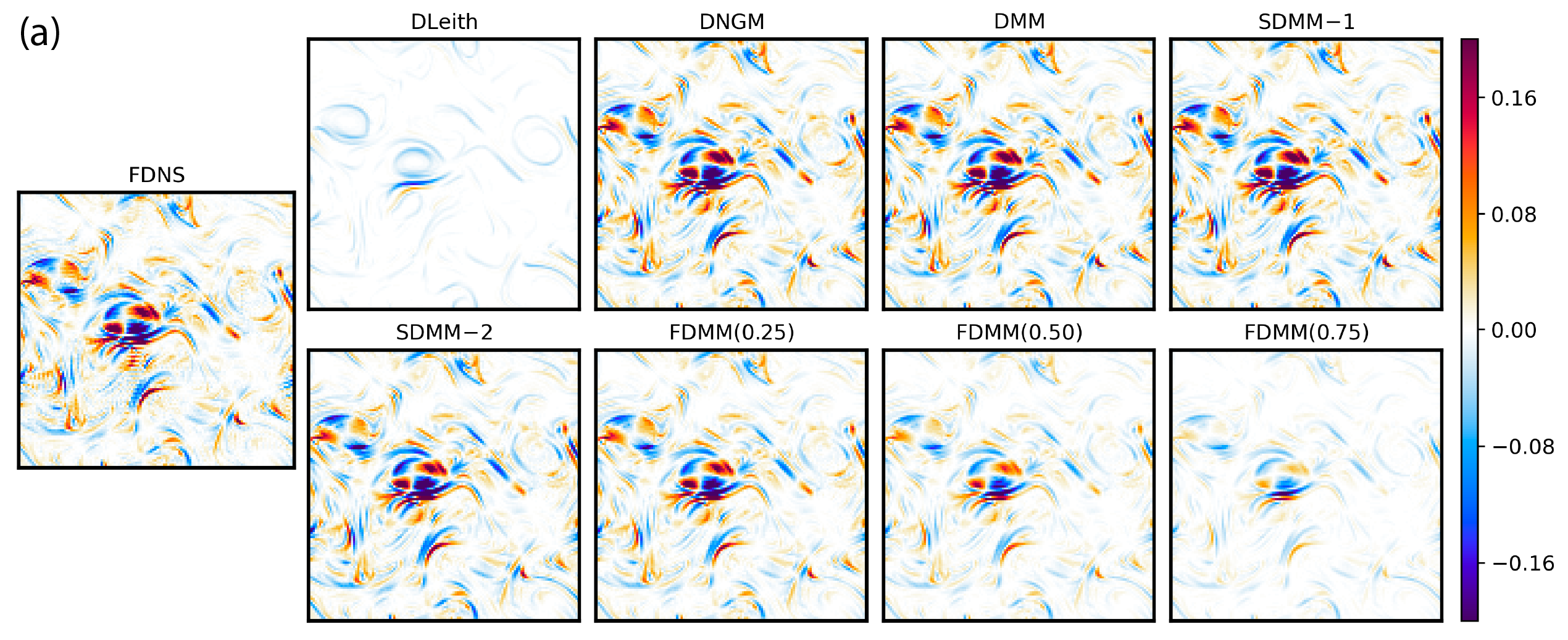}
    \vspace{0.3em}
    \includegraphics[width=1\linewidth]{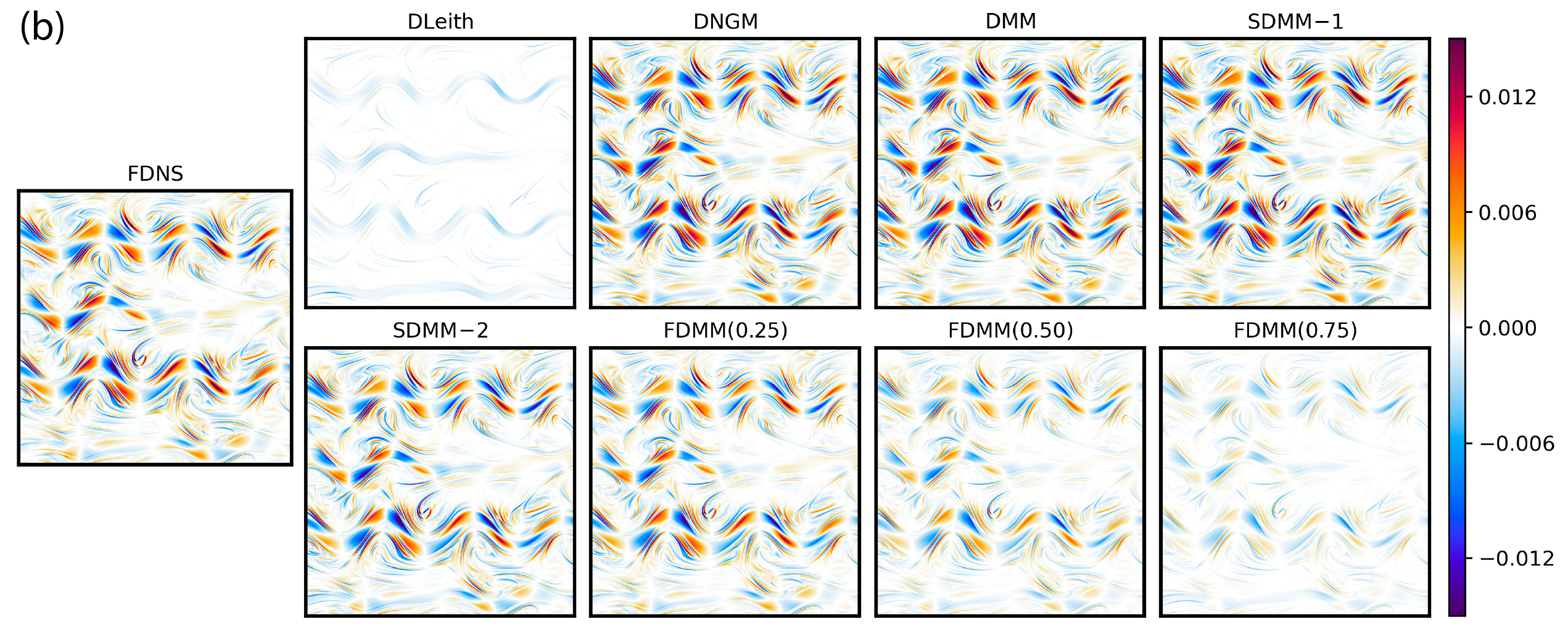}
    \caption{\emph{A priori} tests in the (a) eddy regime (resolution of $128^2$) and (b) jet regime (resolution of $512^2$): Comparison of the instantaneous snapshots of the local SGS exchange of resolved kinetic energy, $\varepsilon^\ell_{\text{sgs}}$. Positive values indicate local backscatter, and negative values indicate local dissipation of resolved kinetic energy.} \label{fig:aprio_sgs_ke_eddy}
\end{figure}

Table\;\ref{tab:aprio_bs} shows the performance of the various models using CC\,($\varepsilon_{\text{sgs}}^{\ell,\,FDNS}, \varepsilon_{\text{sgs}}^{\ell,\,M}$). Trends are broadly consistent with those obtained for $\text{CC}\;(\Pi^{FDNS},\Pi^M)$ (Table\;\ref{tab:aprio_sgs}), although the purely functional DLeith model shows a weaker correlation for local energy exchange. Instantaneous snapshots are shown in  Fig.\;\ref{fig:aprio_sgs_ke_eddy}, corresponding to the same DNS vorticity fields as Fig.\;\ref{fig:test_cases}. Regions of intense backscatter are in proximity to regions of intense dissipation in the ideal exchange from the FDNS \cite{beaumard2026derivation}. Consistent with the weak CC, DLeith is dominated by dissipation and does not show significant backscatter. The modified dynamic mixed models reproduce the alternating dissipative and backscatter regions more faithfully, although the overall amplitude of local backscatter decreases as the relative weight of the functional component increases. We highlight that our modified mixed models still exhibit local backscatter events, unlike structural models with standard clipping \cite{liu1994properties, prakash2022optimal} that eliminate regions of local backscatter.

\begin{table}[h]
\caption{\label{tab:aprio_bs}
\emph{A priori} comparison of correlations of the SGS energy exchange, CC\,($\varepsilon_{\text{sgs}}^{\ell,\,FDNS}, \varepsilon_{\text{sgs}}^{\ell,\,M}$)
}
\begin{ruledtabular}
\begin{tabular}{c|cccccccc}
 & DLeith & DNGM & DMM & SDMM-1 & SDMM-2 & FDMM(0.25) &  FDMM(0.5) & FDMM(0.75)  \\
\colrule
Eddy, $128^2$ & 0.19 & 0.97 & 0.97 & 0.97 & 0.96 & 0.97 & 0.96 & 0.88 \\
Jet, $128^2$ & 0.05 & 0.99 & 0.99 & 0.99 & 0.99 & 0.99 & 0.99 & 0.99 \\
Eddy, $512^2$ & 0.32 & 0.98 & 0.98 & 0.98 & 0.95 & 0.98 & 0.96 & 0.86 \\
Jet, $512^2$ & 0.14 & 0.99 & 0.99 & 0.99 & 0.98 & 0.99 & 0.98 & 0.93
\end{tabular}
\end{ruledtabular}
\end{table}

\begin{table}[h]
\caption{\label{tab:aprio_energy_transfer}
\emph{A priori} comparison of magnitudes of the net SGS energy exchange, $\langle \varepsilon_{\text{sgs}}^{M} \rangle/\langle \varepsilon_{\text{sgs}}^{\,FDNS}\rangle$}
\begin{ruledtabular}
\begin{tabular}{c|cccccccc}
 & DLeith & DNGM & DMM & SDMM-1 & SDMM-2 & FDMM(0.25) &  FDMM(0.5) & FDMM(0.75)  \\
\colrule
Eddy, $128^2$ & -9.20 & 0.54 & 0.31 & 0.34 & -8.73 & -1.89 & -4.33 & -6.77 \\
Jet, $128^2$ & -13.63 & 0.58 & 0.28 & 0.29 & -13.05 & -2.96 & -6.52 & -10.07 \\
Eddy, $512^2$ & -37.14 & 0.77 & -0.32 & -0.09 & -36.52 & -8.70 & -18.18 & -27.66 \\
Jet, $512^2$ & -294.4 & 0.80 & -4.22 & -3.36 & -293.8 & -73.01 & -146.8 & -222.6
\end{tabular}
\end{ruledtabular}
\end{table}

Next, we evaluate the net SGS energy exchange to determine whether the models reproduce both its sign and magnitude. Following \cite{jakhar2026analytical, choi2026data}, we compute $\langle \varepsilon_{\text{sgs}}^{M} \rangle/\langle \varepsilon_{\text{sgs}}^{\,FDNS}\rangle$, as reported in Table\;\ref{tab:aprio_energy_transfer}.  This metric complements the CC by assessing the net magnitude of the modeled SGS energy exchange, rather than only its spatial structure. $\langle \varepsilon_{\mathrm{sgs}}^{\mathrm{FDNS}}\rangle$ is positive for the cases considered, corresponding to net backscatter, and positive values of the ratio indicate that the modeled SGS exchange has the same sign as the ideal exchange. Negative values indicate that the model reverses the sign of the net exchange and thus predicts net dissipation of resolved kinetic energy. 
Table\;\ref{tab:aprio_energy_transfer} confirms that the purely functional DLeith model is strongly dissipative. Interestingly, although the DNGM, DMM and SDMM-1 models showed similar values of CC\,($\varepsilon_{\text{sgs}}^{\ell,\,FDNS}, \varepsilon_{\text{sgs}}^{\ell,\,M}$) in Table\;\ref{tab:aprio_bs}, their net energy exchange differs substantially. In particular, the sequential model, SDMM-1, in which the structural coefficient is obtained first, captures the magnitude more accurately than the fully-coupled DMM. In contrast, SDMM-2 is more dissipative, with this effect being more pronounced when the Smagorinsky model is used as the functional component (not shown). The FDMM($\alpha$) models become increasingly dissipative as $\alpha$ increases. These results show that modifying the coupling in the dynamic procedure can substantially alter the net inter-scale energy exchange, even when the modeled SGS fields have similar spatial correlations with the ideal exchange. As noted earlier, the net magnitude of the SGS forcing decreases with increasing resolution. The large negative ratios observed at $512^2$ resolution reflect the small magnitude of the net FDNS exchange at this finer resolution. Consequently, even moderate net dissipation by the SGS closures is amplified by the normalization.

\begin{figure}[h!]
    \centering
    \includegraphics[width=0.495\linewidth]{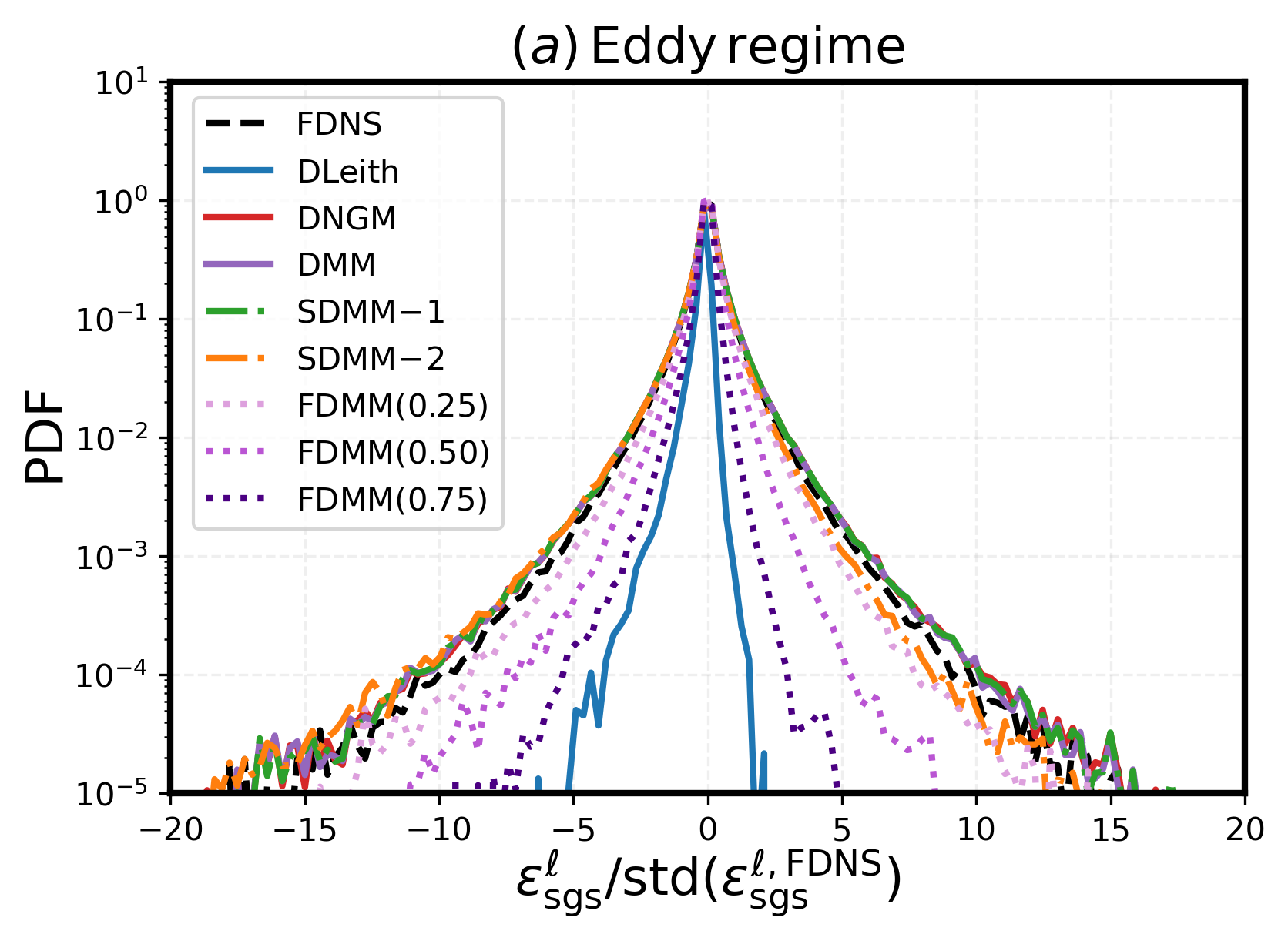}
    \hfill
    \includegraphics[width=0.495\linewidth]{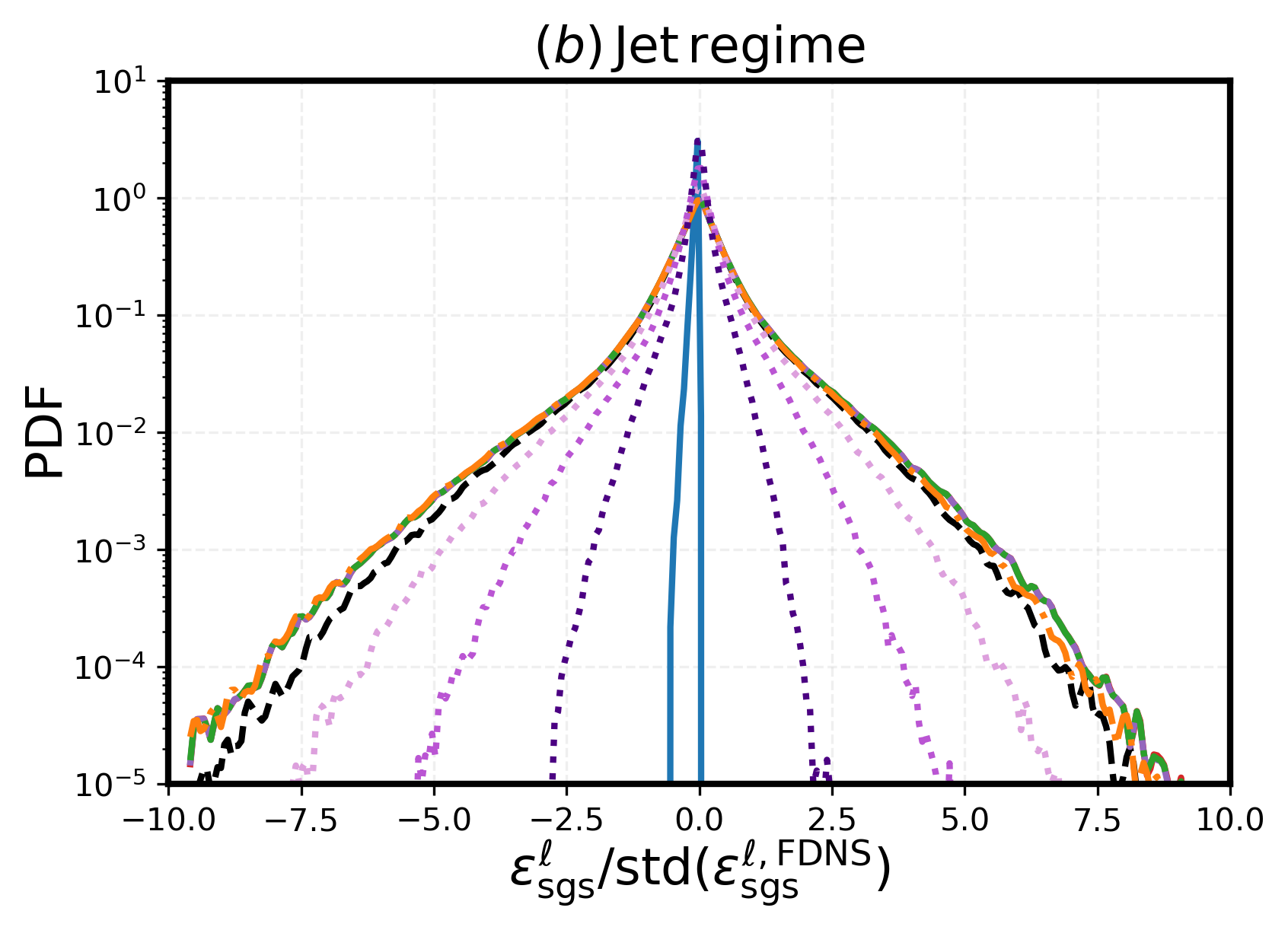}
    \caption{\emph{A priori} tests in the (a) eddy regime and (b) jet regime (both at resolution $128^2$): Probability distribution functions (PDFs) of the subgrid-scale energy exchange $\varepsilon^\ell_{\text{sgs}}$ normalized by the standard deviation (std) of the filtered DNS. Positive values indicate local backscatter, and negative values indicate local dissipation of resolved kinetic energy.}
    \label{fig:aprio_bs_pdf}
\end{figure}

To provide a complementary statistical view of $\varepsilon^\ell_{\text{sgs}}$, we compute its probability distribution function (PDF) using kernel density estimation (KDE), as shown in Fig.\;\ref{fig:aprio_bs_pdf} at a resolution of $128^2$. The FDNS distribution is non-Gaussian with a sharp peak near zero and broad tails, indicating that intense local energy-transfer events occur intermittently \cite{piomelli1991subgrid}.
The distribution is skewed toward dissipative events, although its spatial mean is positive (corresponding to net backscatter) (Table\;\ref{tab:aprio_energy_transfer}), with a larger skewness observed in the eddy regime. The purely functional DLeith model produces a much narrower asymmetric distribution with suppressed tails, strongly underestimating the intermittency and amplitude of local SGS energy-transfer events. In contrast, the structurally dominated models, DNGM, DMM, and SDMM-1, reproduce the broad non-Gaussian tails of the FDNS more closely, demonstrating that the structural component is primarily responsible for capturing intense local backscatter and dissipative events. SDMM-2 also retains the non-Gaussian tails but slightly overestimates the negative tails associated with dissipative events, which could explain its net dissipative bias (Table\;\ref{tab:aprio_energy_transfer}). 

\begin{figure}[h!]
    \centering
    \includegraphics[width=0.95\linewidth]{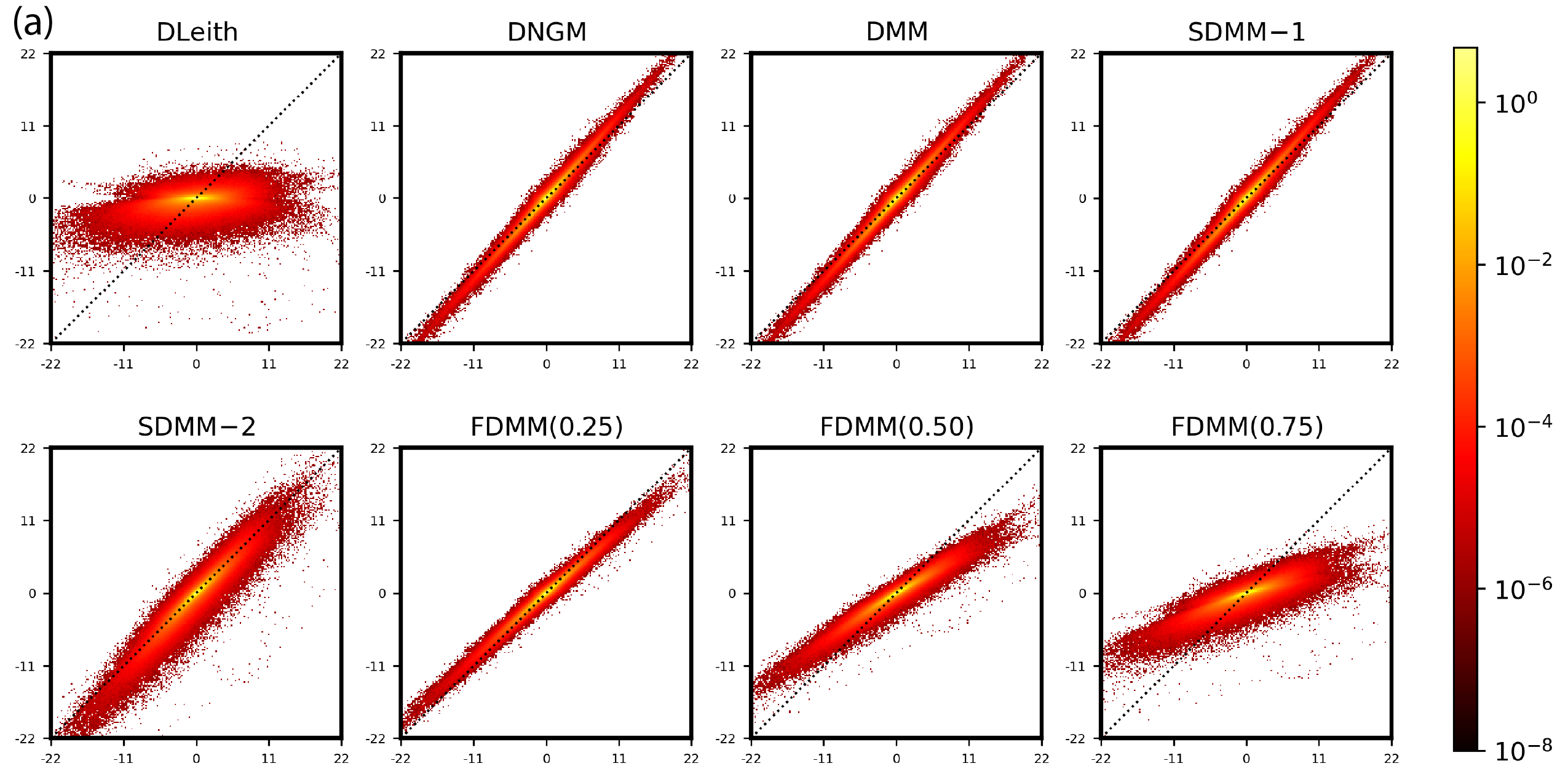}
    \vspace{0.3em}
    \includegraphics[width=0.95\linewidth]{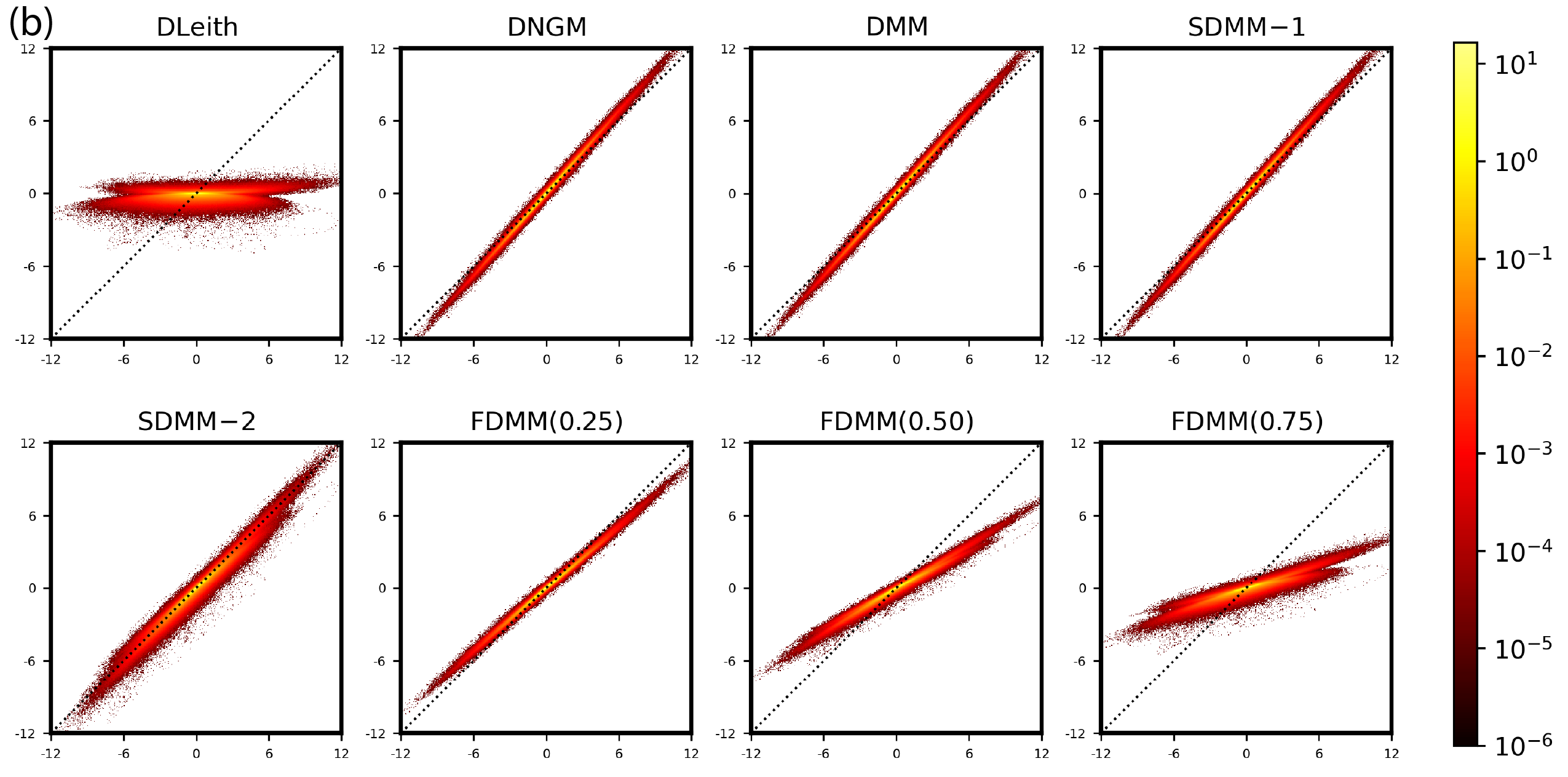}
    \caption{\emph{A priori} tests in the (a) eddy regime and (b) jet regime (both at resolution $512^2$): Joint PDFs of the modeled subgrid-scale energy exchange $\varepsilon^\ell_{\text{sgs}}$ (vertical axis) and its ideal counterpart $\varepsilon^{\ell,\text{FDNS}}_{\text{sgs}}$ (horizontal axis) normalized by the standard deviation (std) of the filtered DNS. The dotted diagonal line highlights deviations from perfect pointwise agreement between the modeled and ideal SGS energy exchange. 
    } 
    \label{fig:aprio_bs_2D_pdf}
\end{figure}

Following \cite{beaumard2026derivation, tuteri2026fluctuations}, we also compute joint PDFs using two-dimensional histograms, as shown in Fig.\;\ref{fig:aprio_bs_2D_pdf} for the $512^2$ resolution. A perfectly modeled SGS would lead to a diagonal joint PDF along the line $\varepsilon_{\mathrm{sgs}}^{\ell, M}=\varepsilon_{\mathrm{sgs}}^{\ell, FDNS}$, indicating a pointwise agreement between the modeled and ideal local energy exchange. Departures from this diagonal indicate deviation in the magnitude or sign of the local SGS energy exchange. For instance, the orientation of the dominant ridge of the joint PDF relative to the diagonal indicates whether the SGS models over- or under-predict the magnitude of the local energy exchange. 
Fig.\;\ref{fig:aprio_bs_2D_pdf} shows that qualitative trends similar to those observed in Fig.\;\ref{fig:aprio_bs_pdf} persist at finer resolution. 
Increasing the relative contribution of the functional model suppresses the tails of the modeled distribution, resulting in horizontally elongated joint PDFs in Fig.\;\ref{fig:aprio_bs_2D_pdf}. The modeled exchange also becomes increasingly skewed toward negative values, indicative of the predominantly dissipative nature of the functional component. This behavior is more pronounced in the jet regime, where the joint PDFs are more tightly concentrated around smaller values of $\varepsilon_{\mathrm{sgs}}^{\ell,M}$ than in the eddy regime. 
Interestingly, the DLeith and other functionally dominated models (FDMM($\alpha$) with $\alpha \to 1$) exhibit weakly bi-lobed joint PDFs, capturing strongly dissipative events but also predicting dissipation in regions with moderate-to-strong FDNS backscatter.
This limitation arises as the Leith model, \eqref{eq:SV_SGS} and \eqref{eq:SV_Leith}, depends solely on local vorticity gradients and not on the alignment with the streamfunction gradients, preventing the Leith model from reliably distinguishing between local forward scatter and backscatter events and thereby resulting in the weak CC in Table\;\ref{tab:aprio_bs}. 

Collectively, the \emph{a priori} tests reveal a wide range in the ability of the modified dynamic mixed model family to reproduce both the spatial structure of the ideal SGS forcing and the net backscatter of energy. The structurally dominated models (DNGM, DMM, and SDMM-1) most accurately reproduce the spatial structure of the SGS forcing and the spatial and statistical distributions of the SGS energy exchange. In contrast, the purely functional model, DLeith, shows low structural fidelity and adds necessary dissipation but does not adequately reproduce backscatter. The sequential model (SDMM-2) retains high structural fidelity but with net dissipation. Across the FDMM($\alpha$) models, a relative increase in the functional component reduces structural fidelity and increases net dissipation.

\subsection{A posteriori tests}

Although \emph{a priori} statistical tests quantify the ability of SGS models to predict instantaneous subgrid forcing and local backscatter using FDNS data, they are insufficient to fully assess the performance of the SGS model. \emph{A posteriori} tests based on fully-coupled LES (i.e., numerical simulations incorporating SGS models) are required to assess the numerical stability and time-integrated accuracy of the resolved fields and their associated energy and enstrophy transfers \cite{pope2001turbulent}. 
Other sources of numerical errors can also significantly influence \emph{a posteriori} tests \cite{sagaut2006large}. In finite difference or finite volume discretizations, numerical dissipation may accumulate and interact with SGS model errors over time, eventually dominating errors \cite{ghosal1996analysis, gullbrand2003effect}. Aliasing errors from the evaluation of nonlinear terms in pseudo-spectral methods can have a similar effect. Our choice of Fourier pseudo-spectral discretization with 2/3-rule dealiasing minimizes these numerical artifacts \cite{jakhar2024learning}, enabling a controlled and fair \emph{a posteriori} evaluation of SGS models.

Due to the chaotic nature of turbulence and 
predictability limit \cite{leith1971atmospheric}, LES trajectories initialized from FDNS fields eventually decorrelate from their FDNS counterparts. Consequently, long-term model performance is quantified using time-averaged statistical quantities. For \emph{a posteriori} tests, we initialize LES with the FDNS vorticity fields well after the flow has reached a statistically self-similar turbulent state. To compute these statistics, LES is time-stepped until $T=60$ (approximately 225 eddy turnover periods) with $dt_\mathrm{LES} = \frac{N_0}{N_\mathrm{LES}}\,dt$ (Sec.\;\ref{sec:test_cases} and App.\;\ref{sec:AppA}).
We analyze the simulated vorticity fields and their non-Gaussian distributions, compensated energy spectrum, and spectral energy and enstrophy transfers.

\subsubsection{Vorticity fields and non-Gaussianity}

\begin{figure}[h]
    \centering
    \includegraphics[width=0.8\linewidth]{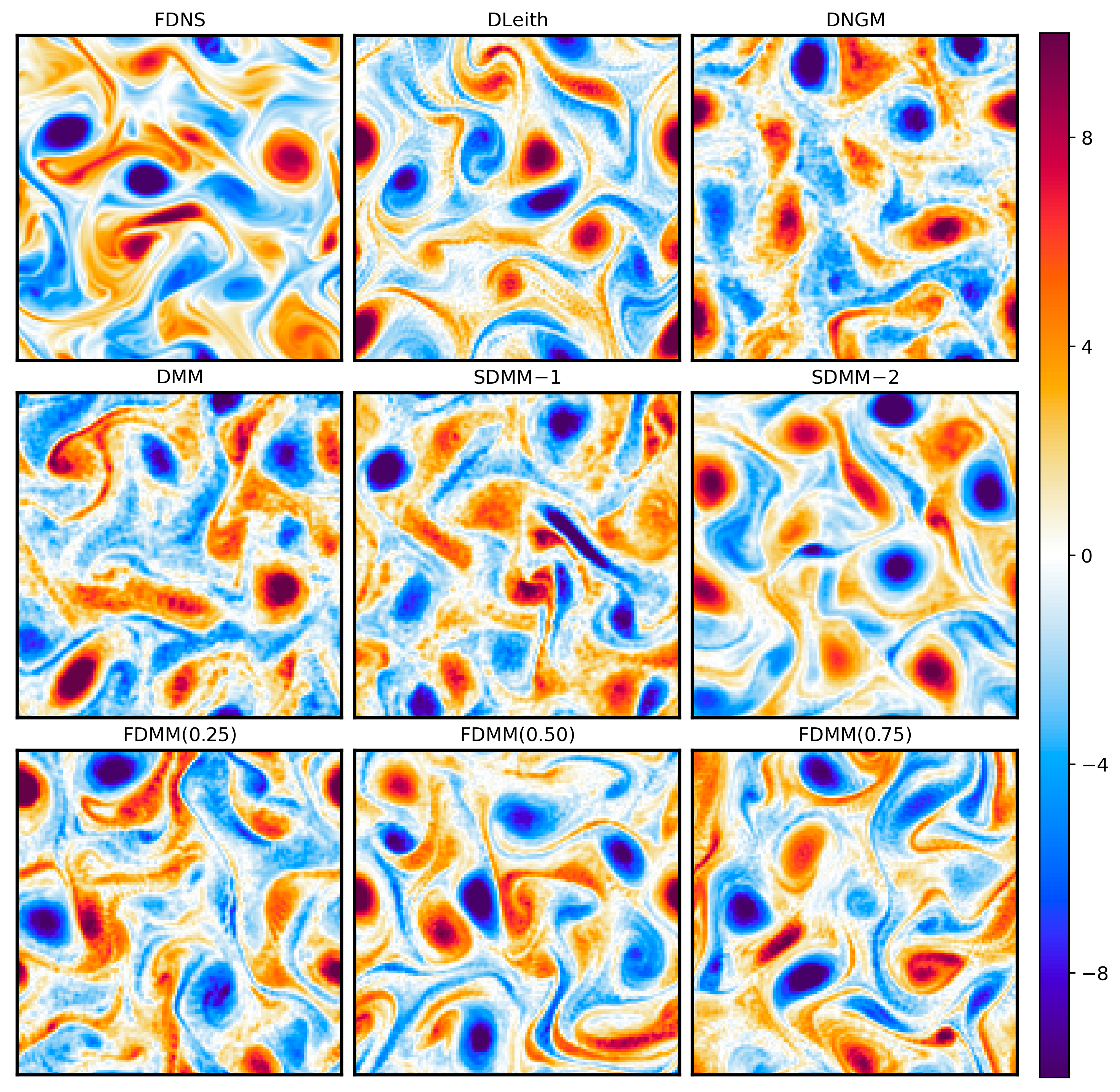}
    \caption{\emph{A posteriori} tests in the eddy regime (resolution of $128^2$): Comparison of instantaneous vorticity snapshots using the modified dynamic mixed models, shown after approximately 225 eddy turnover periods.}
    \label{fig:apost_vort_field}
\end{figure}

We first analyze the qualitative structure of the simulated vorticity fields from LES and the non-Gaussian features of their distributions. Fig.\;\ref{fig:apost_vort_field} shows the instantaneous vorticity fields after approximately 225 eddy turnover periods of LES, in the eddy regime at $128^2$ resolution. All models reproduce large-scale coherent vortices to some extent, with substantial qualitative differences among the models. Thin filamentary structures and regions with high vorticity gradients exhibit noise-like artifacts in the structurally dominated models (DNGM, DMM, SDMM-1). This is likely due to insufficient dissipation during long-term integration, consistent with the net backscatter observed in \emph{a priori} tests (Table\;\ref{tab:aprio_energy_transfer}). The purely functional model (DLeith) suppresses some of these artifacts but introduces fine-scale noise and does not fully recover the FDNS vorticity structure. The FDMM($\alpha$) models show a systematic transition as the functional contribution increases, with better-defined coherent vortical structures but with increased fine-scale noise. In contrast, SDMM-2 provides the closest qualitative agreement with the FDNS field, retaining coherent vortices and fine filamentary structures while suppressing most noise-like artifacts. 

\begin{figure}[h]
    \centering
    \includegraphics[width=0.495\linewidth]{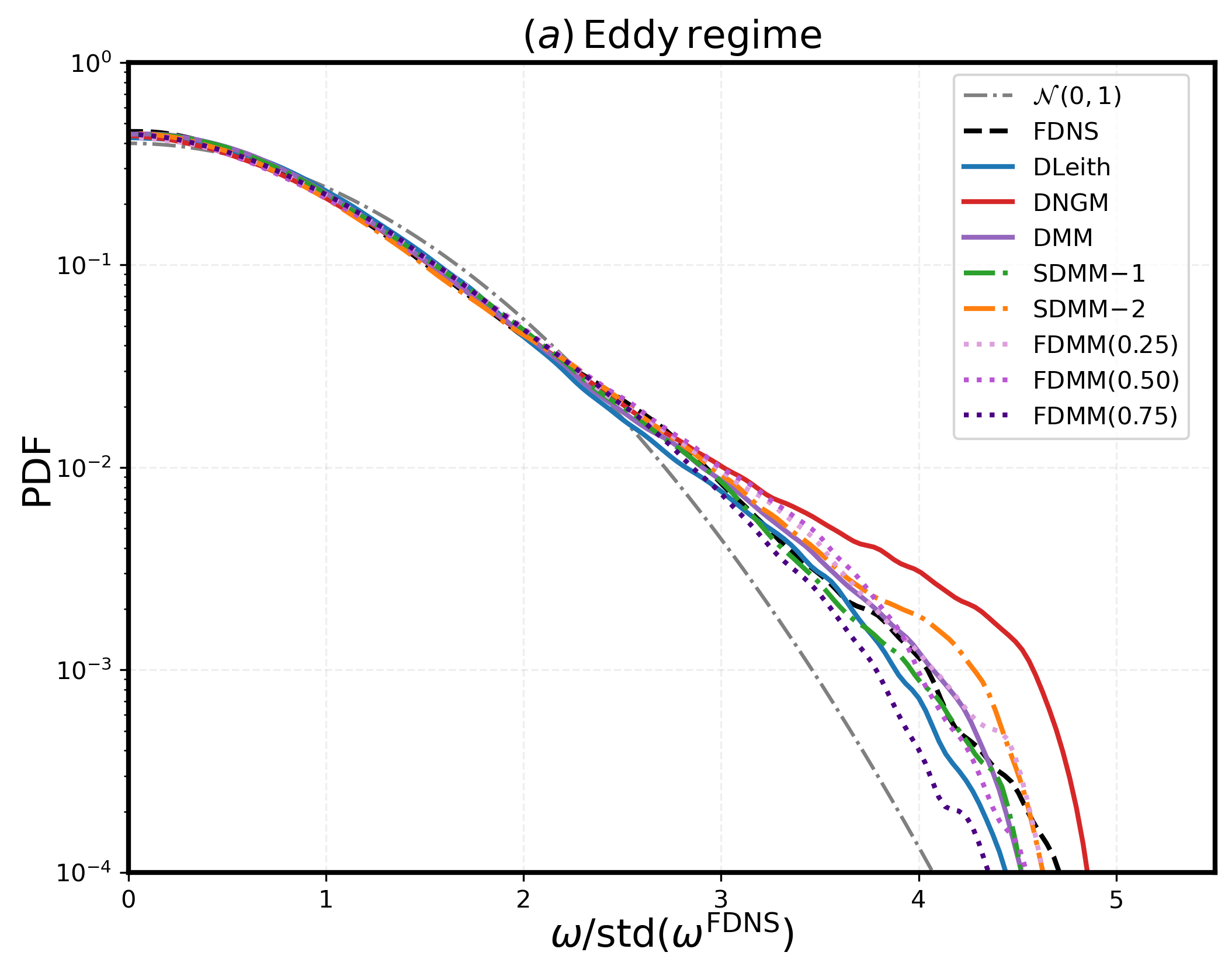}
    \hfill
    \includegraphics[width=0.495\linewidth]{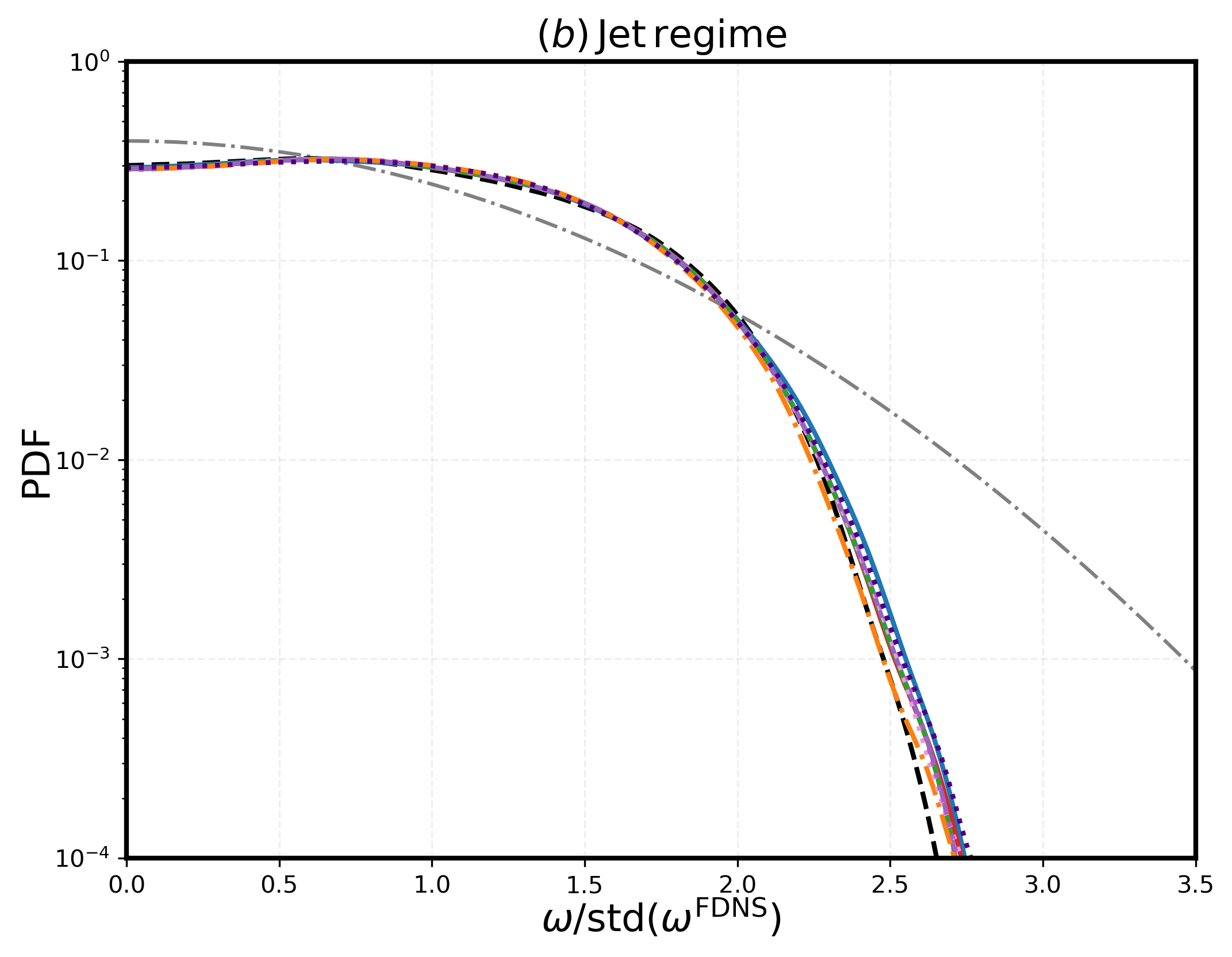}
    \caption{\emph{A posteriori} tests in the (a) eddy regime and (b) jet regime (both at resolution $128^2$): Probability distribution functions (PDFs) of the vorticity fields normalized by the standard deviation (std) of the filtered DNS. A standard normal distribution is overlaid to highlight non-Gaussianity.
    }
    \label{fig:apost_pdf}
\end{figure}

We then assess the non-Gaussianity of the PDFs of vorticity, shown in Fig.\;\ref{fig:apost_pdf}. In the eddy regime, the DNGM overestimates the tails of the distribution, indicating stronger intermittency, whereas the DLeith and the more functionally dominated models produce weaker tails and are less intermittent. In the jet regime, the zonal jets act as mixing barriers; as a result, all models exhibit rapidly decaying, sub-Gaussian tails with comparatively smaller differences among them. We quantify these departures from Gaussianity using higher-order moments, in particular, the kurtosis, $Ku\,(\omega) = \langle \omega^4 \rangle/ \langle \omega^2 \rangle^2$, reported in Table\;\ref{tab:apost_kurt}. A Gaussian distribution has $Ku\,(\omega)=3$ with values above this indicating heavier-than-Gaussian tails and enhanced intermittency, and values below indicating sub-Gaussian behavior. In the jet regime, the organization into zonal jets leads to sub-Gaussian bimodal PDFs \cite{suresh_babu_et_al_JAMES2026} with $Ku\,(\omega)<3$ correctly captured across all models. In the eddy regime, intermittent, long-lived coherent structures dominate the vorticity fields \cite{maltrud1991energy, mcwilliams1999vortices}, resulting in $Ku\,(\omega)>3$. At the coarse resolution of $128^2$, the structurally-dominated models DNGM and DMM best capture kurtosis, while at a resolution of $512^2$, the functional DLeith is best as the filter cut-off is now closer to the dissipative range.

\begin{table}[]
\caption{\label{tab:apost_kurt}
\emph{A posteriori} comparison of the kurtosis of vorticity fields, Ku\;($\omega$)
}
\begin{ruledtabular}
\begin{tabular}{c|ccccccccc}
& FDNS & DLeith & DNGM & DMM & SDMM-1 & SDMM-2 & FDMM(0.25) &  FDMM(0.5) & FDMM(0.75)  \\
\colrule
Eddy, $128^2$ & 4.25 & 3.82 & 4.09 & 4.03 & 3.81 & 3.97 & 3.87 & 3.78 & 3.71 \\
Jet, $128^2$ & 2.10 & 2.09 & 2.07 & 2.06 & 2.07 & 2.06 & 2.08 & 2.08 & 2.08 \\
Eddy, $512^2$ & 3.93 & 3.96 & 3.62 & 3.84 & 4.01 & 4.10 & 3.86 & 3.98 & 4.29 \\
Jet, $512^2$ & 2.12 & 2.09 & 2.10 & 2.10 & 2.10 & 2.10 & 2.11 & 2.10 & 2.10 
\end{tabular}
\end{ruledtabular}
\end{table}

\subsubsection{Long-term evolution of resolved integral quantities}
\begin{figure}[!b]
    \centering
    \includegraphics[width=\linewidth]{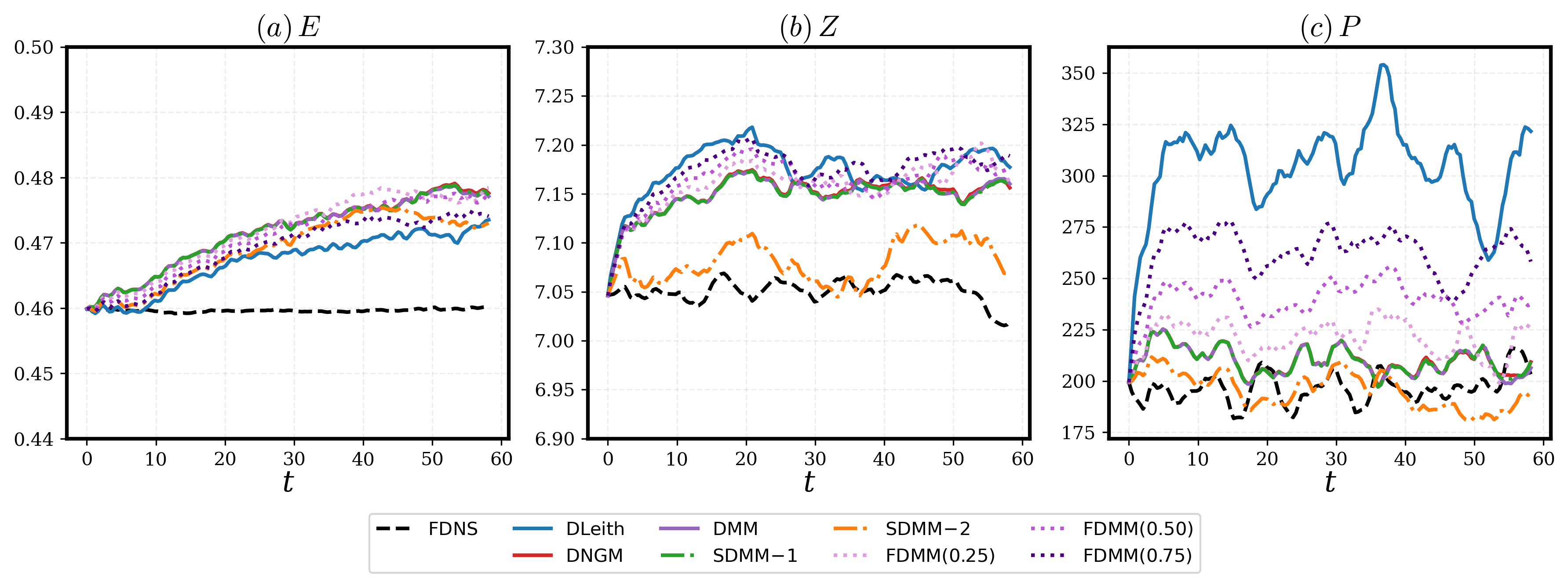}
    \caption{\emph{A posteriori} tests in the jet regime at resolution $128^2$: Temporal evolution of the resolved domain-averaged (a) energy ($E$), (b) enstrophy ($Z$), and (c) palinstrophy ($P$).
    }
    \label{fig:apost_stab}
\end{figure}
To quantify the long-term behavior, we inspect the temporal evolution of the resolved integral quantities: energy, enstrophy, and palinstrophy, following \cite{thuburn2014cascades, frezat2022posteriori}. These quantities provide increasingly scale-selective diagnostics: the kinetic energy is often dominated by the largest resolved scales, the enstrophy highlights the squared vorticity, and the palinstrophy emphasizes high-gradient, high-wavenumber regions.

From Fig.\;\ref{fig:apost_stab}, all models remain bounded and do not exhibit numerical blow-up. The resolved kinetic energy shows a gradual increase relative to the FDNS reference, but with small deviations between models. We note that this slight increase cannot be attributed solely to the net SGS energy exchange, since the overall energy budget also depends on viscous dissipation, dissipation by bottom drag, and energy injection by the forcing \eqref{eq:SGS_budget}. The SGS models lead to variations in the resolved fields (Fig.\;\ref{fig:apost_vort_field}) and therefore affect all the terms of \eqref{eq:SGS_budget_2}, leading to strengthening or weakening of the effects of viscosity, bottom drag, and forcing. Errors arising from temporal discretization can also contribute to the observed discrepancies. 
For enstrophy, SDMM-2 remains comparatively closer to the FDNS reference, with all other models within approximately $5\%$ of FDNS. The largest deviation between models is observed in the palinstrophy. The DLeith model exhibits $\sim\,50\%$  increase in palinstrophy, indicating larger vorticity gradients, followed by the structurally dominated models. Once again, SDMM-2 shows the closest agreement with the FDNS reference. Because palinstrophy is especially sensitive to high-wavenumber vorticity gradients, these deviations are next examined using the compensated energy spectrum. At the finer resolution ($512^2$), all models show much better agreement with the FDNS reference ($< 2\%$ errors, not shown).

\subsubsection{Energy spectrum} 

\begin{figure}[!b]
    \centering
    \includegraphics[width=0.495\linewidth]{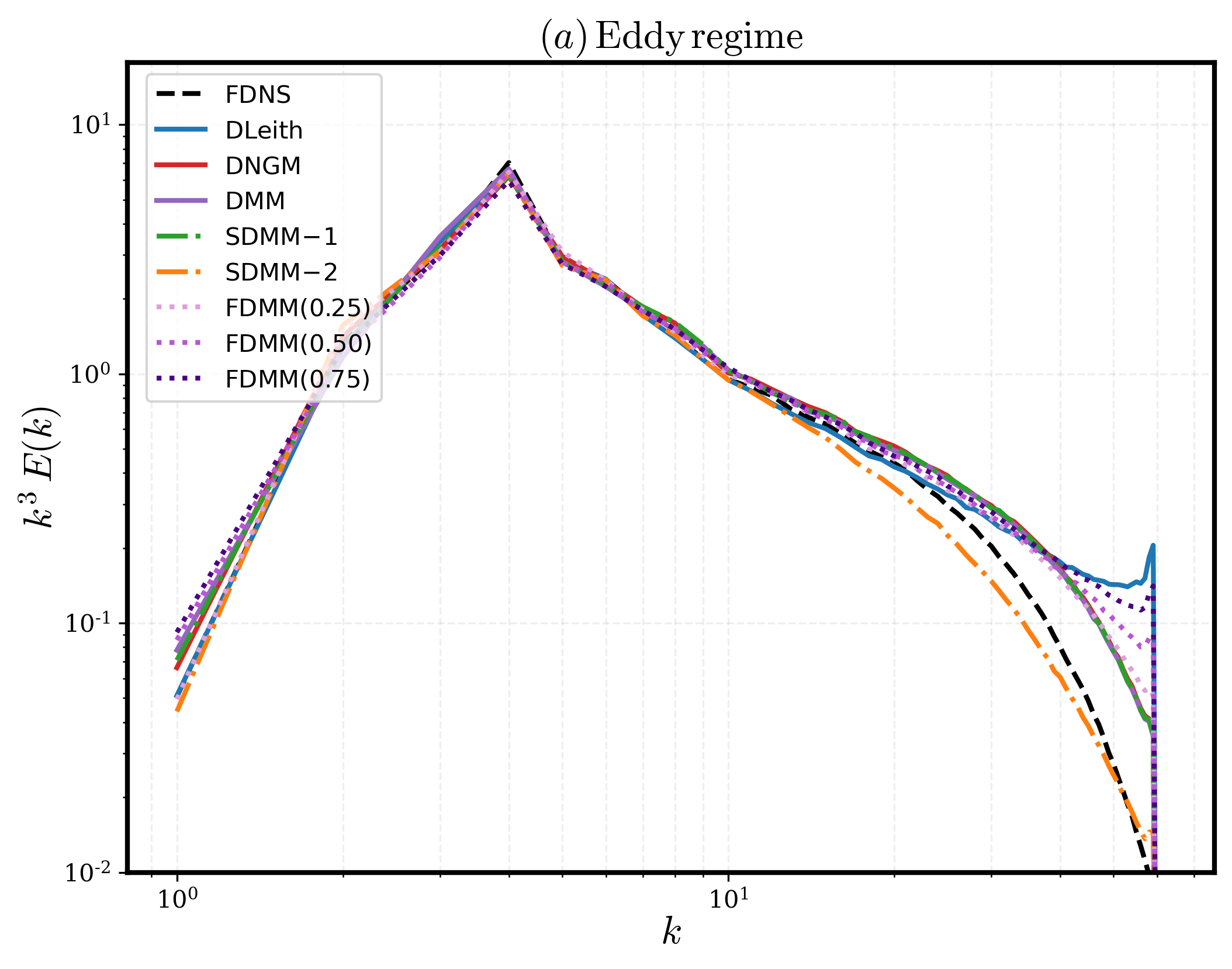}
    \hfill
    \includegraphics[width=0.495\linewidth]{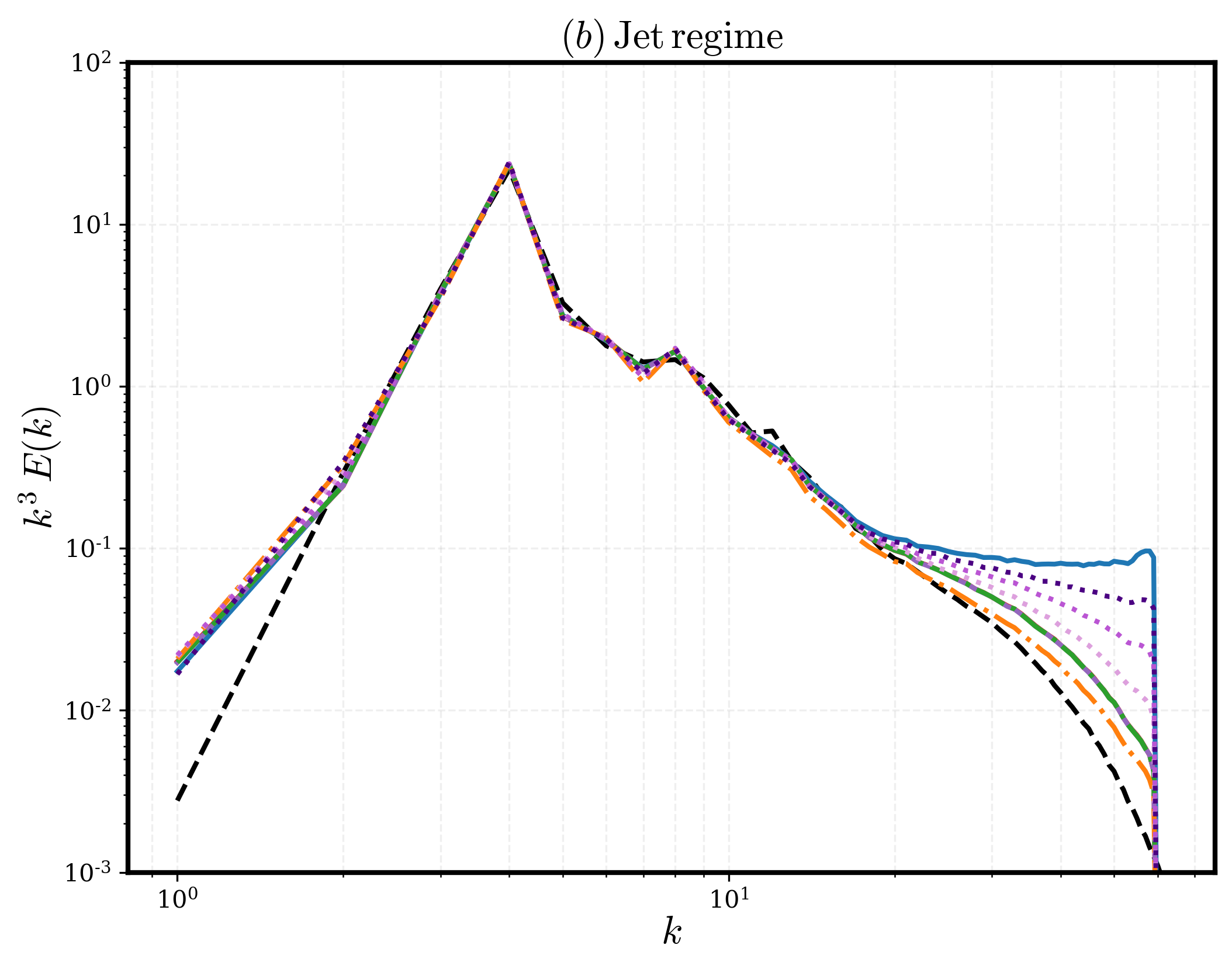}
    \caption{\emph{A posteriori} tests in the (a) eddy regime and (b) jet regime (both at resolution $128^2$): Comparison of time-averaged compensated energy spectra.}
    \label{fig:apost_ke_spec}
\end{figure}

Following \cite{graham2013framework}, model performance across scales is evaluated using the compensated energy spectrum, $k^{3}\,E(k)$ \eqref{eq:E_spectrum}. This compensation emphasizes the high-wavenumber range and hence can help identify numerical instabilities and energy pile-up near the spectral cut-off \cite{boyd2001chebyshev}. At $128^2$ resolution, the sequential model SDMM-2 yields the energy spectrum closest to the FDNS reference (Fig.\;\ref{fig:apost_ke_spec}), followed by the structurally dominated models (DNGM, DMM, and SDMM-1). This agreement is consistent with the smaller palinstrophy deviation observed for SDMM-2 in Fig.\;\ref{fig:apost_stab}, suggesting that SDMM-2 retains much of the strong structural performance from \emph{a priori} tests (Table\;\ref{tab:aprio_sgs}), with the functional component providing sufficient dissipation to stabilize the time integration and suppress high-wavenumber energy pile-up. In contrast, structurally dominated models reproduce the SGS forcing and energy transfers with high accuracy \emph{a priori} (Table\;\ref{tab:aprio_energy_transfer}), but provide less dissipative regularization in long-term simulations, consistent with instantaneous vorticity fields (Fig.\;\ref{fig:apost_vort_field}). 
This illustrates the well-known mismatch between \emph{a priori} accuracy and \emph{a posteriori} performance \cite{park2005toward}. While dynamically obtained closures may provide instantaneous SGS forcing with high structural correlation, this does not guarantee stable long-term evolution of vorticity fields. Small errors can accumulate and alter the resolved trajectory, particularly when the model exhibits excessive backscatter or provides insufficient local dissipation.
Although DLeith and the other functionally dominated models are predominantly dissipative in the \emph{a priori} analysis 
and provide a net downscale transfer, they under-predict the magnitude and intermittency of the SGS energy exchange. This is consistent with the narrow tails of their local SGS energy-exchange ($\varepsilon^\ell_{\text{sgs}}$) PDFs (Fig.\;\ref{fig:aprio_bs_pdf}), which indicate that large intermittent transfer events are not adequately represented. As a result, their local dissipation is insufficient during long-term integration, leading to energy accumulation at high wavenumbers despite the model being dissipative on average. Similar spectral behavior for dynamic subgrid-viscosity models was reported in \cite{jakhar2026analytical}. At $512^2$ resolution (not shown), the structurally dominated models capture the FDNS energy spectrum, including the spectral tails, while the DLeith shows improved performance but still exhibits slight deviations near the spectral cut-off.

\subsubsection{Spectral energy and enstrophy transfers}
\begin{figure}[!b]
    \centering
    \includegraphics[width=1\linewidth]{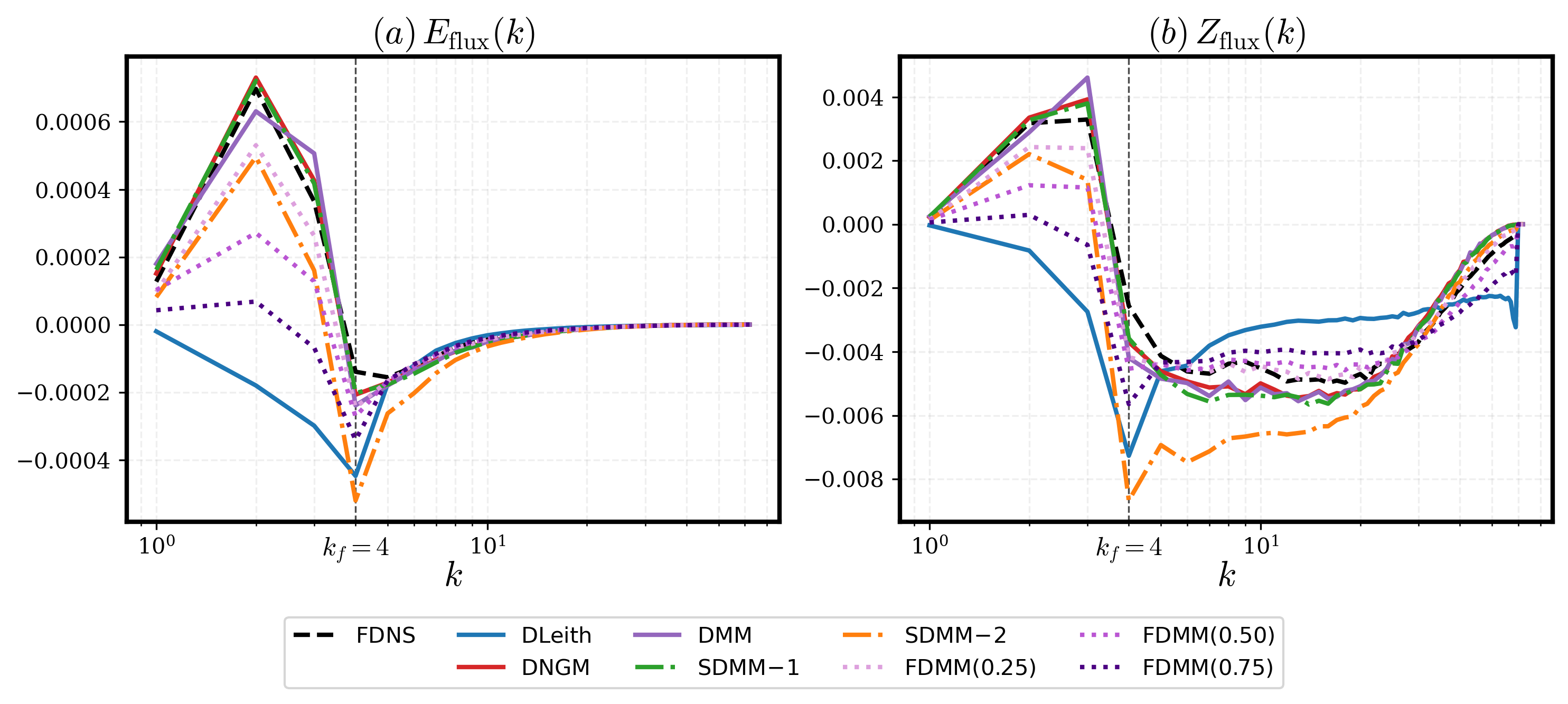}
    \vspace{0.3em}
    \includegraphics[width=1\linewidth]{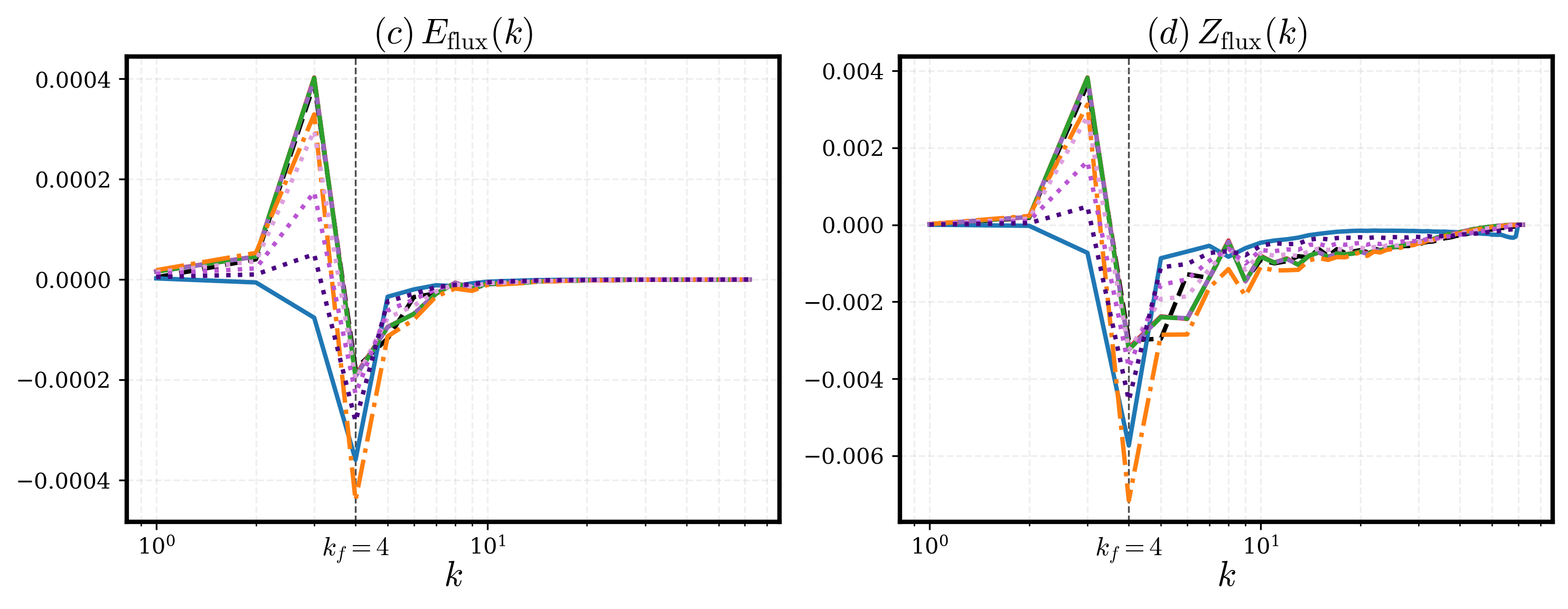}
    \caption{\emph{A posteriori} tests in the (a, b) eddy regime and (c, d) jet regime (all at resolution $128^2$): Comparison of time-averaged spectral energy and enstrophy transfers. Positive values indicate backscatter, and negative values indicate dissipation. A vertical line is overlaid to indicate the forcing wavenumber.} 
    \label{fig:apost_pi_ezk}
\end{figure}

Finally, to explain the spectral deviations observed, the ability of the SGS models to reproduce the spectral energy and enstrophy transfers is examined. Complementary to the physical-space SGS energy-exchange diagnostics, the spectral transfer functions quantify the contribution of the SGS forcing to the resolved energy and enstrophy budgets at each wavenumber shell $k$. As in \cite{maltrud1993energy, srinivasan2024turbulence}, we define the spectral energy and enstrophy transfers (inter-scale fluxes) as
\begin{equation} \label{eq:EZ_flux}
E_{flux}(k) = \oiint_{|\mathbf{k}|=k} \Re(-\Pi(\mathbf{k})^*\;\Tilde{\fpsi}(\mathbf{k}))
\, dS, \qquad
Z_{flux}(k) = \oiint_{|\mathbf{k}|=k}
\Re(\Pi(\mathbf{k})^*\;\Tilde{\fomega}(\mathbf{k}))\, dS,
\end{equation}
where $\Re({\bullet})$ indicates the real part, and $\bullet^{*}$ is the complex conjugate. Hence, negative values of $E_{flux}(k)$ and $Z_{flux}(k)$ indicate net dissipation from resolved to subgrid-scales, while positive values indicate backscatter. 

The FDNS reference shows significant backscatter at wavenumbers smaller than the forcing wavenumber $k_f$ (i.e., $k < k_f$), corresponding to scales larger than the forcing scale, and dissipation at larger wavenumbers ($k > k_f$) (Fig.\;\ref{fig:apost_pi_ezk}). 
Consistent with the \emph{a priori} analysis (Fig.\;\ref{fig:aprio_sgs_ke_eddy}), the DLeith model is predominantly dissipative across all wavenumbers and therefore does not reproduce the large-scale backscatter. Meanwhile, the structurally dominated models (DNGM, DMM, SDMM-1) reproduce the sign and magnitude of the reference spectral transfers more accurately, including backscatter, although their enstrophy transfers show slight deviations at intermediate-to-large wavenumbers. This behavior is consistent with their high structural fidelity but weaker regularizing effect in long-term simulations. SDMM-2 produces a more balanced transfer. Notably, it retains the backscatter at large scales, with particularly good agreement in the jet regime, while producing enhanced dissipation at wavenumbers slightly larger than the forcing wavenumber $k_f$. This additional forward transfer could explain why SDMM-2 yields a more regularized vorticity field and a compensated energy spectrum closer to FDNS at $128^2$ resolution. At $512^2$ resolution (not shown), this trend persists; the SDMM-2 model exhibits larger dissipation but still retains some large-scale backscatter. In contrast, the FDMM(0.75) and DLeith models remain dissipative across all wavenumbers.

Cases in which $\delta_{12}$ and $\delta_{21}$ are allowed to vary continuously are summarized in Appendix\;\ref{sec:AppC}. Results show that these models interpolate between SDMM-1 and SDMM-2.
Overall, the \emph{a posteriori} tests confirm that the functional-structural coupling of the modified dynamic mixed models determines the trade-off between instantaneous fidelity and long-term stability and accuracy.

\section{\label{sec:conclusions} Conclusions}

We constructed modified dynamic mixed subgrid-scale (SGS) models for idealized geophysical flows such as forced two-dimensional and beta-plane turbulence. By analyzing the functional and structural components' contributions to the Germano identity error (GIE), we showed that the least-squares coefficient estimation in a fully-coupled dynamic mixed model (DMM) is biased toward the structural component. To overcome this structural dominance, we decomposed the Gram matrix into self- and cross-interaction terms and developed a modified system of equations that enables the control of the coupling between the functional and structural components. The resulting parametric family of modified mixed models includes two sequential formulations (SDMM-1 and SDMM-2), in which one component is dynamically estimated first and the other acts as a correction, and fully-decoupled formulations (FDMM($\alpha$)), in which the relative influence of the functional and structural components is explicitly controlled through the mixing parameter $\alpha$.

To examine how the functional-structural coupling affects the trade-off between \emph{a priori} (instantaneous) fidelity and \emph{a posteriori} (time-integrated) accuracy, we then used an idealized mesoscale beta-plane framework and combined the Leith model with the fourth-order nonlinear gradient model (NGM4). \emph{A priori} tests at two LES resolutions confirmed that the functional Leith model (DLeith) shows weak correlation (CC $\approx 0.3 - 0.6$) with the ideal SGS forcing and does not account for the upscale transfer of energy from unresolved to resolved scales (backscatter). In contrast, the structurally dominated models (DNGM, DMM, SDMM-1) show strong correlation (CC $> 0.97$) and reproduce the spatial distribution and the heavy non-Gaussian tails of the local SGS exchange, with substantially better representation of net backscatter, particularly at coarse resolution. A sequential model in which the functional component is determined first and then corrected by the structural component (SDMM-2) retains strong correlation and local energy exchange distributions while introducing larger net dissipation. The FDMM($\alpha$) models further confirm that increasing the functional contribution monotonically reduces structural fidelity while enhancing dissipation, providing a controllable interpolation between the structural and functional models.

\emph{A posteriori} tests show that these \emph{a priori} trends do not translate directly into time-integrated LES accuracy. The structurally dominated models (DNGM, DMM, SDMM-1) lead to noisy vorticity gradients and thin filamentary structures with large high-wavenumber deviations in the compensated spectrum, indicating insufficient net dissipation and reduced regularization from the functional component, while DLeith exhibits significant fine-scale noise due to insufficient local dissipation. SDMM-2 leads to the closest qualitative agreement with the filtered DNS vorticity fields and spectra, accompanied by close agreement with domain-averaged quantities such as energy, enstrophy, and palinstrophy. Analysis of the spectral energy and enstrophy transfers reveals that this sequential model exhibits scale-selective dissipation, retaining much of the backscatter at large scales (wavenumbers smaller than the forcing wavenumber), with enhanced dissipation at smaller scales, leading to a larger regularizing effect during long-term simulations.

While the present work is limited to idealized two-dimensional and beta-plane turbulence, future studies could extend the modified mixed modeling framework to incorporate active or passive tracer transport \cite{gent1995parameterizing}, stratification \cite{ozgokmen2007large, ozgokmen2011large}, and more realistic oceanic gyres and coastal domains \cite{greatbatch2000four, suresh_babu_et_al_Oceans2025}. The role of localized averaging based on Lagrangian pathlines and flow maps can also be investigated \cite{meneveau1996lagrangian, kulkarni_lermusiaux_JCP2019}. Incorporating non-Markovian closures, for example, based on Mori-Zwanzig formalism \cite{parish2017non, gupta_lermusiaux_PRSA2021}, and stochastic parameterizations based on \emph{optimal} LES formulations \cite{langford1999optimal} would also be useful, and could be combined with data-driven closures \cite{gupta_lermusiaux_SR2023, ross2023benchmarking, ivagnes2023hybrid,srinivasan2024turbulence}.

\appendix
\section{\label{sec:AppA}
 Definition of filtering operators}

While the choice of numerical grid and discretization schemes can implicitly filter out unresolved high-frequency scales in LES, additional explicit filters, such as the box, sharp spectral cut-off, or Gaussian filters, can also be applied \cite{sagaut2006large}. In this work, we consider explicit homogeneous filters $G$ that act on the full field $\phi$ to produce the filtered field $\overline{\phi}$. In physical space, we have,
\begin{equation}
    \overline{\phi}(\mathbf{x}) = \int G(\mathbf{x}-\mathbf{x'}) \phi(\mathbf{x'}) d\mathbf{x'},
\end{equation}
or equivalently, in spectral space,
\begin{equation}
    \Tilde{\overline{\phi}}(\mathbf{k}) = \Tilde{G}(\mathbf{k})\,\Tilde{\phi}(\mathbf{k}).
\end{equation}
where $\Tilde{ \bullet}$ is the Fourier transform, $\Tilde{G}(\mathbf{k})$ is the transfer function of the filtering operator,  $\mathbf{k}=(k_x,k_y)$ is the horizontal wavenumber vector, and $k=|\mathbf{k}|=\sqrt{k_x^2+k_y^2}$ is the horizontal angular wavenumber.

In particular, we obtain filtered variables using a Gaussian filter, followed by coarse-graining to the LES grid using a sharp spectral cut-off with transfer functions
\begin{equation}\label{eq:filters}
        \Tilde{G}_{\text{Gaussian}}(k) = \exp\bigg(\frac{-k^2 \Delta_F^2}{24}\bigg), \qquad \Tilde{G}_{\text{cut-off}}(k) = \begin{cases}
        1, \qquad \text{if}\,\,k \leq k_c\\
        0, \qquad \text{otherwise}
        \end{cases},
\end{equation}
where $\Delta_F$ is the filter width and $k_c$ is the cut-off wavenumber. The same transfer functions \eqref{eq:filters} are used for test filtering, with $\widehat{\Delta}_\text{F} = 2\,\Delta_\text{F}$.

For LES with resolution $N^2_{LES}$, we utilize $\Delta_F =  2 \Delta_{LES}$ where $\Delta_{LES} = \frac{2 \pi}{N_{LES}}$, and $k_c = \frac{ \pi}{\Delta_{LES}} = \frac{N_{LES}}{2}$ \cite{pope2001turbulent}.

\section{\label{sec:AppB} Temporal evolution equations for energy and enstrophy}

Starting from \eqref{eq:2D_EZP}, an alternate form for the kinetic energy can be derived 
from Green's first identity \cite{quartapelle2013numerical}, 
\begin{equation}\label{eq:appB_green}
    \int \phi \nabla^2 \chi \,dA  = - \int \nabla \phi \cdot \nabla \chi \,dA + \oint \phi \nabla \chi \cdot\mathbf{n}\,ds,
\end{equation}
where $\oint(\bullet).\mathbf{n}\,ds$ denotes a contour integral with outward normal $\mathbf{n}$. This contour integral vanishes under periodic boundary conditions, and hence we obtain the following expressions for the kinetic energy and its time evolution equation using $\omega = \nabla^2 \psi$, 
\begin{eqnarray} \label{eq:appB_balance}
    \begin{gathered}
         E= \frac{1}{2} \langle | \nabla \psi |^2 \rangle = -\frac{1}{2} \langle  {\psi} \omega \rangle, \\
         \frac{d E}{d t} = \frac{1}{2}\frac{d \langle | \nabla \psi |^2 \rangle}{dt} = -\frac{1}{2}\frac{d \langle \psi \omega \rangle}{dt}  = \frac{1}{2} \bigg[ -\left\langle \frac{d \psi}{dt} \omega \right \rangle - \left\langle\psi \frac{d \omega}{dt}  \right \rangle\bigg ].
     \end{gathered}
\end{eqnarray}

Repeating the same process, we obtain
\begin{equation}\label{eq:appB_approx}
       \left \langle \frac{d \psi}{dt} \omega\right\rangle =\left \langle \frac{d \psi}{dt} \nabla^2 \psi\right\rangle  = -\left \langle \frac{d(\nabla \psi)}{dt} \cdot \nabla \psi\right\rangle = -\frac{1}{2}\frac{d\left \langle | \nabla \psi |^2\right\rangle}{dt}.
\end{equation}

Substituting \eqref{eq:appB_approx} in \eqref{eq:appB_balance} yields
\begin{equation}
    \frac{d E}{dt} = \frac{1}{2} \frac{d E}{dt} - \frac{1}{2}\left\langle\psi \frac{d \omega}{dt} \right\rangle,
\end{equation}
\begin{equation}
    \implies \frac{d E}{dt} = -\left\langle\psi \frac{d \omega}{dt}\right\rangle.
\end{equation}

The time evolution equation for enstrophy can be obtained more directly using the chain rule,
\begin{eqnarray} \label{eq:appB_enstrophy}
        Z= \frac{1}{2} \langle \omega ^2 \rangle, \qquad
         \frac{d Z}{d t} =\left\langle\omega\frac{d \omega}{dt}\right\rangle.
\end{eqnarray}
\section{\label{sec:AppC} Sequential dynamic mixed models with continuously varying modifiers}

We summarize the \emph{a priori} and \emph{a posteriori} performance of sequential dynamic mixed models \eqref{eq:Gmod_linsys} with modifiers that take continuous values between zero and one. SDMM($\kappa$) denotes the model obtained by setting $\delta_{12} = \kappa$ and $\delta_{21} = 1 - \kappa$ with $\kappa \in \{0.25,0.50,0.75\}.$ These values of $\kappa$ linearly interpolate the modifiers between the two limiting sequential formulations SDMM-1 ($\delta_{12} = 0,\, \delta_{21}=1$) and SDMM-2 ($\delta_{12} = 1,\, \delta_{21}=0$).

In \emph{a priori} tests (not shown), the SDMM($\kappa$) models yield values of $c_{sv}$ identical to those of the corresponding FDMM($\alpha$) models ($\alpha \approx \kappa$), but with much larger values of $c_{ngm}$ that lie between those of SDMM-1 and SDMM-2. Other \emph{a priori} metrics of the SDMM($\kappa$) models also interpolate between SDMM-1 and SDMM-2. In particular, they yield correlations of $\Pi$ and $\varepsilon_{\text{sgs}}^{\ell}$ between those of the two limiting sequential models (Tables\;\ref{tab:aprio_sgs} and \ref{tab:aprio_bs}), with an increase in $\kappa$ resulting in lower structural fidelity. The magnitudes of their net SGS energy exchange are comparable to those obtained with the corresponding FDMM($\alpha$) models (Table\;\ref{tab:aprio_energy_transfer}).

\begin{figure}[!h]
    \centering
    \includegraphics[width=0.495\linewidth]{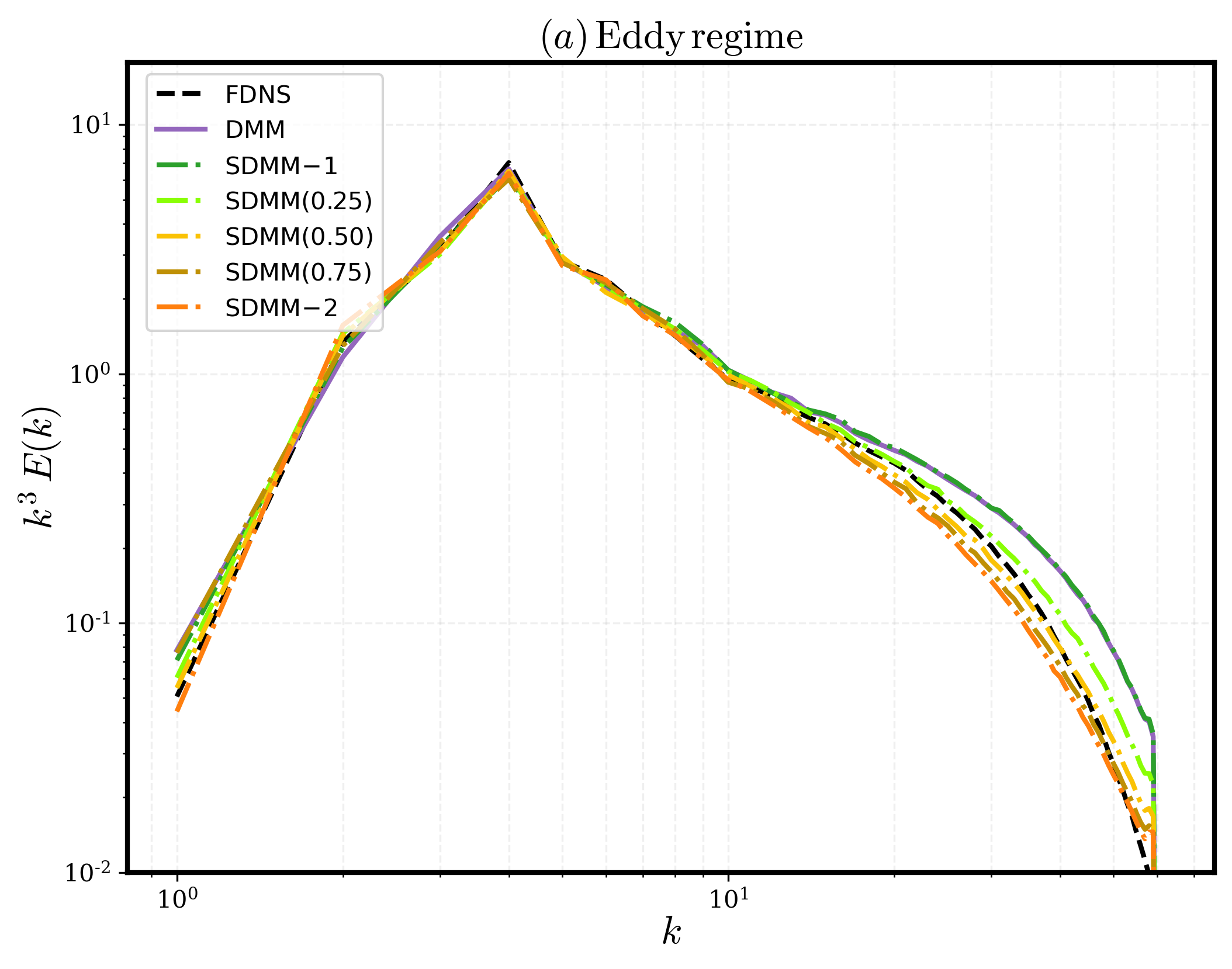}
    \hfill
    \includegraphics[width=0.495\linewidth]{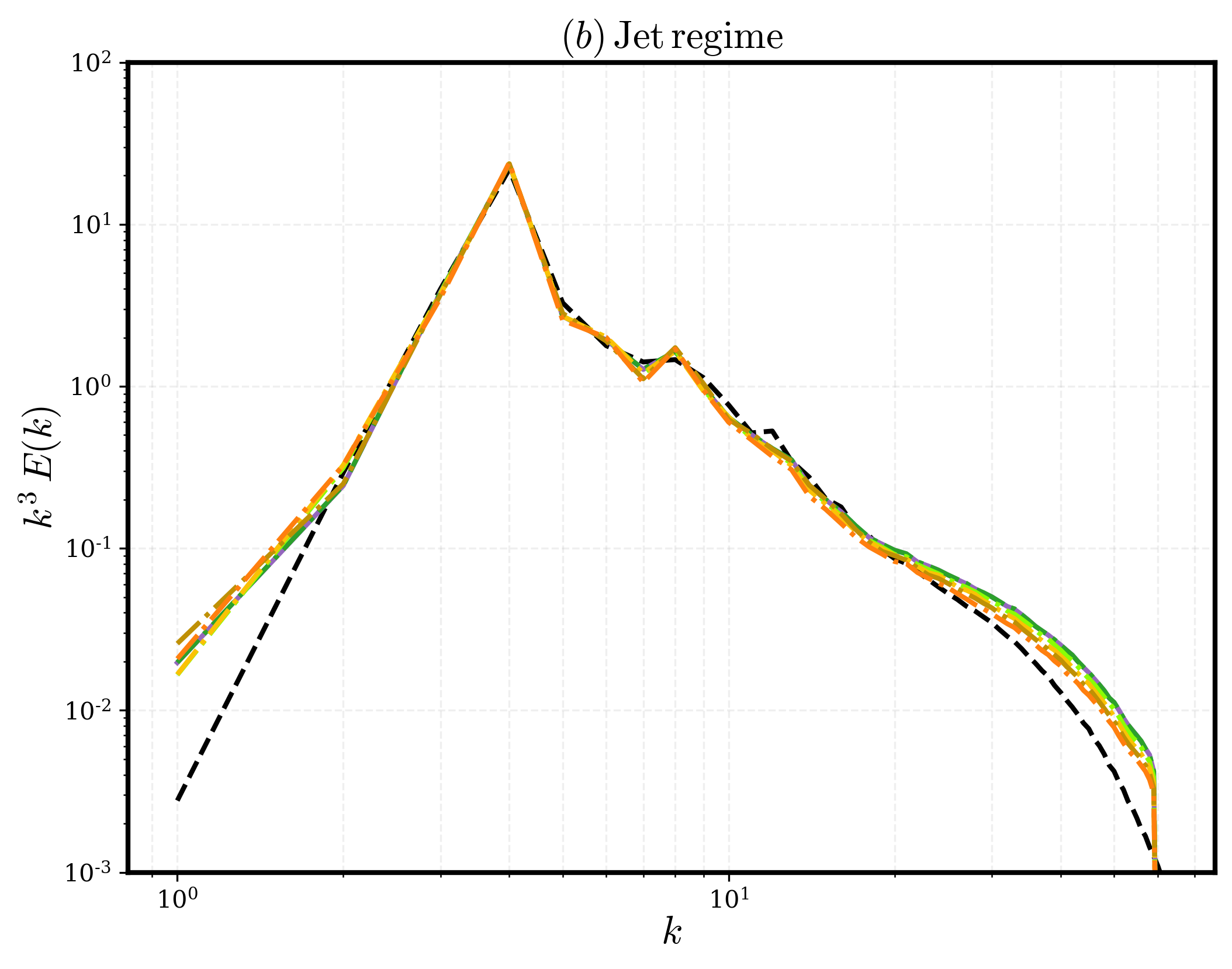}
    \caption{As Fig.\;\ref{fig:apost_ke_spec}, but for sequential dynamic mixed models with continuously varying modifiers.}
    \label{fig:app_ke_spec}
\end{figure}

In \emph{a posteriori} tests, decreasing $\kappa$ leads to more pronounced noise-like artifacts. As shown in Fig.\;\ref{fig:app_ke_spec}, the kinetic energy spectra of the SDMM($\kappa$) models again lie between those of the two limiting sequential formulations. Similar behavior is observed in the temporal evolution of domain-averaged quantities and in the spectral energy and enstrophy transfers. However, the SDMM($\kappa$) models show much weaker forward transfer compared to SDMM-2 at wavenumbers just above the forcing wavenumber $k_f$, consistent with larger values of $c_{ngm}$ in \emph{a priori} tests.
%

\begin{acknowledgments}
We thank the members of our MSEAS group, especially A.A.\ Pophale for proofreading the manuscript, and the ML-SCOPE MURI team for helpful discussions. We are grateful to the Office of Naval Research for partial support under grant N00014-20-1-2023 (MURI ML-SCOPE) and N00014-24-1-2715 (DRI-RIOT) to the Massachusetts Institute of Technology. 
\end{acknowledgments}

\bibliography{mseas, intro, methods, results}

\end{document}